\documentclass[a4paper]{article}
\usepackage{jheppub}
\usepackage{lineno}
\usepackage{hyperref}


\newcommand{\pt}{\ensuremath{p_{\rm T}}}

\newcommand{\pythia}{{\sc pythia}}
\newcommand{\powheg}{{\sc powheg}}
\newcommand{\powhegbox}{{\sc powheg-box}}
\newcommand{\powhegpyt}{{\sc powheg+pythia}}
\newcommand{\powhegher}{{\sc powheg+herwig}}
\newcommand{\herwig}{{\sc herwig}}
\newcommand{\jimmy}{{\sc jimmy}}
\newcommand{\herwigpp}{{\sc herwig{\small ++}}}
\newcommand{\alpgen}{{\sc alpgen}}
\newcommand{\hej}{{\sc hej}}
\newcommand{\geant}{{\sc geant4}}

\title{Measurement of dijet production with a veto on additional central jet activity in $pp$ collisions at $\sqrt{s}=7$~TeV using the ATLAS detector}

\usepackage{subfigure}

\author{The ATLAS Collaboration}

\abstract{
A measurement of jet activity in the rapidity interval bounded by a dijet system is presented. Events are vetoed if a jet with transverse momentum greater than 20 GeV is found between the two boundary jets.
The fraction of dijet events that survive the jet veto is presented for boundary jets that are separated by up to six units of rapidity and with mean transverse momentum $50 < \overline{p}_{\rm T} < 500$~GeV. 
The mean multiplicity of jets above the veto scale in the rapidity interval bounded by the dijet system is also presented as an alternative method for quantifying perturbative QCD emission. 
The data are compared to a next-to-leading order plus parton shower prediction from the \powhegbox{}, an all-order resummation using the \hej{} calculation and the \pythia, \herwigpp{} and \alpgen{} event generators. The measurement was performed using $pp$ collisions at $\sqrt{s}=7$~TeV using data recorded by the ATLAS detector in 2010. 
}

\proceeding{CERN-PH-EP-2011-100}

\begin{document}

\maketitle 

\section{Introduction\label{seq:intro}}
Dijet production with a veto on additional hadronic activity in the rapidity interval between the jets 
has previously been 
studied at HERA \cite{Derrick:1995pb, Adloff:2002em, Chekanov:2006pw} and the Tevatron 
\cite{Abachi:1994hb,Abe:1994de, Abe:1998b,Abbott:1998jb, Abe:1998a}. The Large Hadron Collider 
(LHC) offers the opportunity to study this process at an increased 
centre-of-mass energy and with a wider coverage in rapidity between jets. Historically, the main purpose of these measurements has been to 
search for evidence of colour singlet exchange. With this aim, a very low cut on the
total hadronic activity between the jets (less than a few GeV) was traditionally 
chosen, to suppress contributions from colour octet exchange. 

In this measurement, a jet veto is used to identify the absence of additional activity.
This approach is useful because it allows a diverse range of perturbative QCD phenomena to be studied, as the veto scale is chosen to be much larger than $\Lambda_{\rm QCD}$. First, BFKL-like dynamics\footnote{BFKL dynamics propose an evolution in $\ln(1/x)$, where $x$ is the
Bjorken variable, as opposed to the DGLAP \cite{Altarelli:1977zs} evolution in $\ln(Q^2)$, where $Q^2$ is the parton virtuality} \cite{Fadin:1975cb,Kuraev:1976ge,Kuraev:1977fs,Balitsky:1978ic} are expected to become increasingly important for large rapidity intervals \cite{Andersen:2009nu,Andersen:2010ih,Forshaw:2005sx,Mueller:1986ey}. Alternatively, the effects of wide-angle soft-gluon radiation can be studied in the limit that the average dijet transverse momentum is much larger than the scale used to veto on additional jet activity \cite{Forshaw:2009fz,Forshaw:2006fk}. Finally, colour singlet exchange is expected to be important if both limits are satisfied at the same time, i.e the jets are widely separated and the jet veto scale is small in comparison to the dijet transverse momentum. The measurement is therefore targeted at studying the effects of QCD radiation in those regions of phase space that may not be adequately described by standard event generators. 

A central jet veto is also used in the search for Higgs production via vector boson fusion in the Higgs-plus-two-jet channel in order to reject backgrounds. Furthermore, should the Higgs boson be discovered, the contribution from gluon fusion to this channel needs to be determined in order to extract the Higgs boson couplings \cite{Duhrssen:2004cv,Lafaye:2009vr,Campbell:2006xx,Cox:2010ug}. This measurement, therefore, could be used to constrain the theoretical modelling in current Higgs searches and future precision Higgs measurements.

\section{The ATLAS detector}
ATLAS is a general-purpose detector surrounding interaction point one of
the LHC \cite{detPaper, ATLAS}. The main detector components relevant to this analysis are the inner tracking detector, the calorimeters and the minimum bias trigger scintillators (MBTS). 
The inner tracking detector covers the pseudorapidity range $|\eta|<2.5$, and has full coverage in azimuth.\footnote{ATLAS uses a right-handed coordinate system with its origin at the nominal interaction point (IP) in the centre of the detector and the $z$-axis along the beam pipe. The $x$-axis points from the IP to the centre of the LHC ring, and the $y$ axis points upward. Cylindrical coordinates $(r,\phi)$ are used in the transverse plane, $\phi$ being the azimuthal angle around the beam pipe. The pseudorapidity is defined in terms of the polar angle $\theta$ as $\eta=-\ln\tan(\theta/2)$. The rapidity of a particle with respect to the beam axis is defined
as $y = \frac{1}{2}\ln\frac{E+p_z}{E-p_z}$.} 
There are three main components to the inner tracker; the silicon pixel detector, the silicon microstrip detector and the transition-radiation detector. These components are arranged in concentric layers and immersed in a 2~T magnetic field provided by the inner solenoid magnet. 

The ATLAS calorimeter is also divided into sub-detectors. The electromagnetic calorimeter ($|\eta|
< 3.2$) is a high-granularity sampling detector in which the active medium is liquid argon (LAr) inter-spaced with layers of lead absorber. The hadronic calorimeters are divided into three sections: a tile scintillator/steel calorimeter is used in both the barrel ($|\eta|<1.0$) and extended barrel cylinders ($0.8 < |\eta| < 1.7$); the hadronic endcap covers the region $1.5 < |\eta| < 3.2$ and consists of LAr/copper calorimeter modules; the forward calorimeter measures both electromagnetic and hadronic energy in the range $3.2 < |\eta| < 4.9$ using LAr/copper and LAr/tungsten modules. The total coverage of the ATLAS calorimeters is $|\eta|<4.9$. 

The primary triggers used to readout the ATLAS detector were the calorimeter jet triggers \cite{trignote}. The calorimeter jet triggers were validated for this measurement using a fully efficient minimum bias trigger derived from the MBTS. The MBTS consists of 32 scintillator counters arranged on two disks located in front of the end-cap calorimeter cryostats. The MBTS cover the region $2.1 < |\eta| < 3.8$.

\section{Measurement definition}\label{sec:measdef}
Jets are reconstructed using the anti-$k_{\rm t}$ algorithm \cite{Cacciari:2008gp} with distance parameter $R=0.6$ and full four momentum recombination. Jets are required to have transverse momentum $p_{\rm T}^{}>20$~GeV and rapidity $|y| <4.4$, ensuring that they are in a region in which the jet energy scale has been validated (Sec.~\ref{sec:jes}). The dijet system is identified using two different selection criteria. In the first, the two highest transverse momentum jets in the event are used, which probes wide-angle soft gluon radiation in $p_{\rm T}$-ordered jet configurations. In the second, the most forward and the most backward jets in the event are used (i.e. those with the largest rapidity separation, $\Delta y$), which favours BFKL-like dynamics because the dijet invariant mass is much larger than the transverse momentum of the jets. For both definitions, the mean transverse momentum of the jets that define the dijet system, $\overline{p}_{\rm T}$, is required to be greater than 50~GeV. This ensures that the measurement is in the high efficiency region of the calorimeter jet trigger (Sec.~\ref{sec:evsel}). 

Two variables are used to quantify the amount of additional radiation in the rapidity interval bounded by the dijet system. The first is the gap fraction, which is the fraction of events that do not have an additional jet with a transverse momentum greater than a given veto scale, $Q_{0}$, in the rapidity interval bounded by the dijet system. The default value of the veto scale is chosen to be $Q_{0}=20$~GeV. The second variable is the mean number of jets with $p^{}_{\rm T}>Q_{0}$ in the rapidity interval bounded by the dijet system. The measurements of these two variables are fully corrected for experimental effects. The final distributions therefore correspond to the `hadron-level', in which the jets are reconstructed using all final state particles that have a proper lifetime longer than 10~ps. This includes muons and neutrinos.

\section{Monte Carlo event simulation}
Simulated proton-proton collisions
at $\sqrt{s} = $ 7 TeV were produced using Monte Carlo (MC) event 
generators. These samples were used to derive systematic uncertainties and
correct for detector effects. The reference generator 
was \pythia{} 6.4.2.3 \cite{pythia}, which implements leading-order (LO) QCD matrix elements for $2\rightarrow 2$ processes followed by a $p_{\rm T}$-ordered parton shower and  the Lund string model of hadronisation. The underlying event in 
\pythia{} is modelled by multiple parton interactions interleaved with the initial 
state parton shower. The events were generated using the MRST LO* parton distribution functions (PDF)  \cite{Martin:2009iq,Sherstnev:2008dm} and the AMBT1 tune \cite{ATLAS-CONF-2010-031}. The final state particles were passed through a detailed \geant{} \cite{geant4} simulation of the ATLAS detector \cite{AtlasSimu} and reconstructed using the same analysis chain as for the data.  

Fully simulated event samples were also generated using \herwigpp{} 2.5.0 \cite{Bahr:2008pv} and \alpgen{} \cite{ref:noteALPGEN}. \herwigpp{} implements leading order $2\rightarrow 2$ matrix elements, but uses an angular-ordered parton shower and a cluster hadronisation model. The underlying event is modelled by multiple parton interactions. The \herwigpp{} event samples are generated using the MRST LO* PDF set with the LHC-UE7-1 tune for the underlying event \cite{Gieseke:2011na}.   
\alpgen{} provides LO matrix elements with up to six partons in the final state. The \alpgen{} samples are generated using the CTEQ6L1 PDF set  \cite{Pumplin:2002vw}
and passed through \herwig{} 6.5 \cite{herwig} and \jimmy{} \cite{Butterworth:1996zw} 
to provide parton showering, hadronisation and multiple partonic interactions with tune AUET1  \cite{auet}. 

\section{Theory predictions} 
The measurements presented in this paper probe perturbative QCD in the region where the energy scale of the dijet system is larger than the scale of the additional radiation. At large values of $\overline{p}_{\rm T}/Q_{0}$ or $\Delta y$, it is expected that fixed order calculations are unlikely to describe the data and that a resummation to all orders in perturbation theory is necessary. The measurement is not particularly sensitive to non-perturbative physics because $Q_0$ is chosen to be much greater than $\Lambda_{\rm QCD}$. 
The net effect of the non-perturbative physics corrections was estimated by turning the hadronisation and underlying event on and off in \pythia{} -- the resulting shift in the gap fraction was less than 2\% and the change in the mean number of jets in the rapidity interval bounded by the dijet system was less than 4\%. 

The theoretical predictions were produced using \hej{}  \cite{Andersen:2010ih,Andersen:2011hs} and the \powhegbox{} \cite{Nason:2004rx,Alioli:2010xd,Alioli:2010xa}.  
\hej{} is a parton-level event generator that provides an all-order description  
of wide-angle emissions of similar transverse momentum.  In this BFKL-inspired limit, \hej{} reproduces the full QCD results and is especially suited for events with at least two jets separated by a large rapidity interval\footnote{In the default setup (used in this analysis), \hej{} matches the resummation to leading order $2\rightarrow3$ and $2\rightarrow4$ matrix elements. However, the option to include the additional running coupling terms from next-to-leading-log BFKL was not used.}. The events were generated with the MSTW 2008 NLO PDF set \cite{Martin:2009iq} and the partons were clustered into jets using the anti-$k_{\rm t}$ algorithm with distance parameter $R=0.6$.  The renormalisation/factorisation scale (one parameter in \hej{}) was chosen to be the $p_{\rm T}^{}$ of the leading parton and the uncertainty due to this choice was estimated by increasing and decreasing the scale by a factor of two. The uncertainty from the choice of PDF was estimated using the full set of eigenvector errors provided by MSTW and also by changing the PDF to CTEQ61 \cite{Pumplin:2002vw}. 
The overall uncertainty in the \hej{} calculation is dominated by the scale choice and is typically 5\% for the gap fraction and 8\% for the mean number of jets in the rapidity interval bounded by the dijet system. These uncertainties are larger than the non-perturbative physics corrections and the \hej{} parton-level predictions are therefore used for data-theory comparisons.

The \powhegbox{} provides a full next-to-leading order (NLO) dijet calculation and is interfaced to \pythia{} or \herwig{} to provide all-order resummation of soft and collinear emissions using the parton shower approximation.  
The \powheg{} events were generated using the MSTW 2008 NLO PDF set with the renormalisation and factorisation scales set to the $p_{\rm T}^{}$ of the leading parton. These events were passed through both \pythia{} (tune AMBT1) and \herwig{} (tune AUET1) to provide different hadron-level predictions. The difference between these two predictions was found to be larger than the intrinsic uncertainty in the NLO calculation, estimated by varying the PDFs and the renormalisation and factorisation scales. Therefore, the \powhegpyt{} and \powhegher{} predictions are both used for data-theory comparisons.

\section{Jet reconstruction and energy scale determination}\label{sec:jes}
Jets are reconstructed at detector level using electromagnetic (EM) scale topological clusters\footnote{The electromagnetic scale is the 
basic calorimeter signal scale for the ATLAS calorimeters. It gives
the correct response for the energy deposited in electromagnetic
showers, while it does not correct for the lower hadron response.}, which are three-dimensional objects built from calorimeter cells \cite{topoclusters}. The jet energies are corrected using \pt{} and $\eta$ dependent jet energy scale (JES) calibration factors derived from simulated MC events \cite{calibnote3}.  
The JES calibration is obtained by dividing the true jet energy, defined using stable interacting particles in the MC event record (i.e. excluding muons and neutrinos), by the EM scale energy of the matching detector-level jet. The corrections are derived for jets
with $p_{\rm T}^{}>10$~GeV at the EM scale and parameterised as a function of jet $p_{\rm T}^{}$ and $|\eta|$. An additional correction factor is applied to the $\eta$ of jets that fall in the crack-regions of the detector, to remove the bias caused by the constituents of jets falling in regions of very different calorimeter response. The final stage recalculates the jet kinematics using the primary vertex position, rather than the ATLAS geometric centre (0,0,0). 

The absolute JES uncertainty has been determined using data for the well-understood barrel region ($|\eta|<0.8$), by propagating the uncertainty in the single-particle response, measured by the tracking and calorimeter systems, to the jet constituents \cite{singleparticle}. An additional uncertainty has been obtained for other calorimeter regions using dijet $\eta$-intercalibration \cite{insitu2}; the jet calorimeter response relative to the barrel region was studied by balancing the transverse momenta of dijets and the uncertainty estimated by comparing the results obtained with data to a variety of MC based predictions. The total JES uncertainty in each calorimeter region was taken to be the sum in quadrature of the absolute uncertainty (from the barrel) and relative uncertainty from the dijet intercalibration \cite{calibnote3}.  The final JES uncertainty is approximately 2--5\% in the barrel region, but rises to 13\% in the forward calorimeter for jets with $p_{\rm T}\sim 20$~GeV. 

The impact of the JES uncertainty on the measurement of dijet production with a jet veto was studied by varying the energy scale of jets within the JES uncertainty, allowing for different calorimeter regions to have correlated/uncorrelated calorimeter responses. The associated uncertainty on the gap fraction is typically 3\% (7\%) for $\Delta y \sim 3$~(6). The uncertainty on the mean number of jets in the rapidity interval bounded by the dijet system is approximately 5\% and only weakly dependent on $\Delta y$. The effect of the JES uncertainty is the largest systematic uncertainty in the measurement for most of the phase space regions that are presented.

\section{Event selection}\label{sec:evsel}

The measurement was performed using data taken during 2010. The primary trigger selections used to readout the ATLAS detector were the calorimeter jet triggers. In particular, distinct regions of $\overline{p}_{\rm T}$ were defined and, in each region, only events that passed a specific jet trigger (at least one jet above a defined threshold) were used. The $\overline{p}_{\rm T}$ regions were defined using \pythia{} events and validated using data collected with a minimum bias trigger derived from the MBTS. The chosen trigger for each region was required to be the highest threshold (and therefore least prescaled) trigger that was at least 99\% efficient for both gap events and inclusive events. The bias in the measurement from the use of jet triggers was estimated to be less than 0.25\%. This was determined using MC simulations and also data collected with the minimum bias trigger.

To minimise the impact of pile-up, each event was required to have exactly one reconstructed primary vertex, defined as a vertex with at least five tracks that was consistent with the beamspot. 
The fraction of events in each run with only one reconstructed vertex was 90\% in the early low luminosity runs, falling to 20\% in the high luminosity runs at the end of 2010. With the single vertex selection applied, the gap fraction was observed to be independent of the data taking period 
 and the systematic bias due to residual pile-up events was determined to be less than 0.5\%. 
Events were also rejected if they contained any `fake' jets with 
$p_{\rm T}>20$~GeV that originated from calorimeter noise bursts, cosmic rays or beam-backgrounds. The criteria for rejection were determined from events with spuriously large missing transverse energy. The efficiency for jets was determined to be greater than 99\% using a tag-and-probe method in dijet events. The overall impact of these cleaning cuts was to reduce the number of events by less than 0.4\%. Finally, the impact of beam related backgrounds and cosmic rays were studied using data collected with special-purpose trigger selections, and estimated to be less than 0.1\%.

\begin{figure}[t]
\centering
\mbox{
  \subfigure[]{ \includegraphics[width=.32\textwidth]{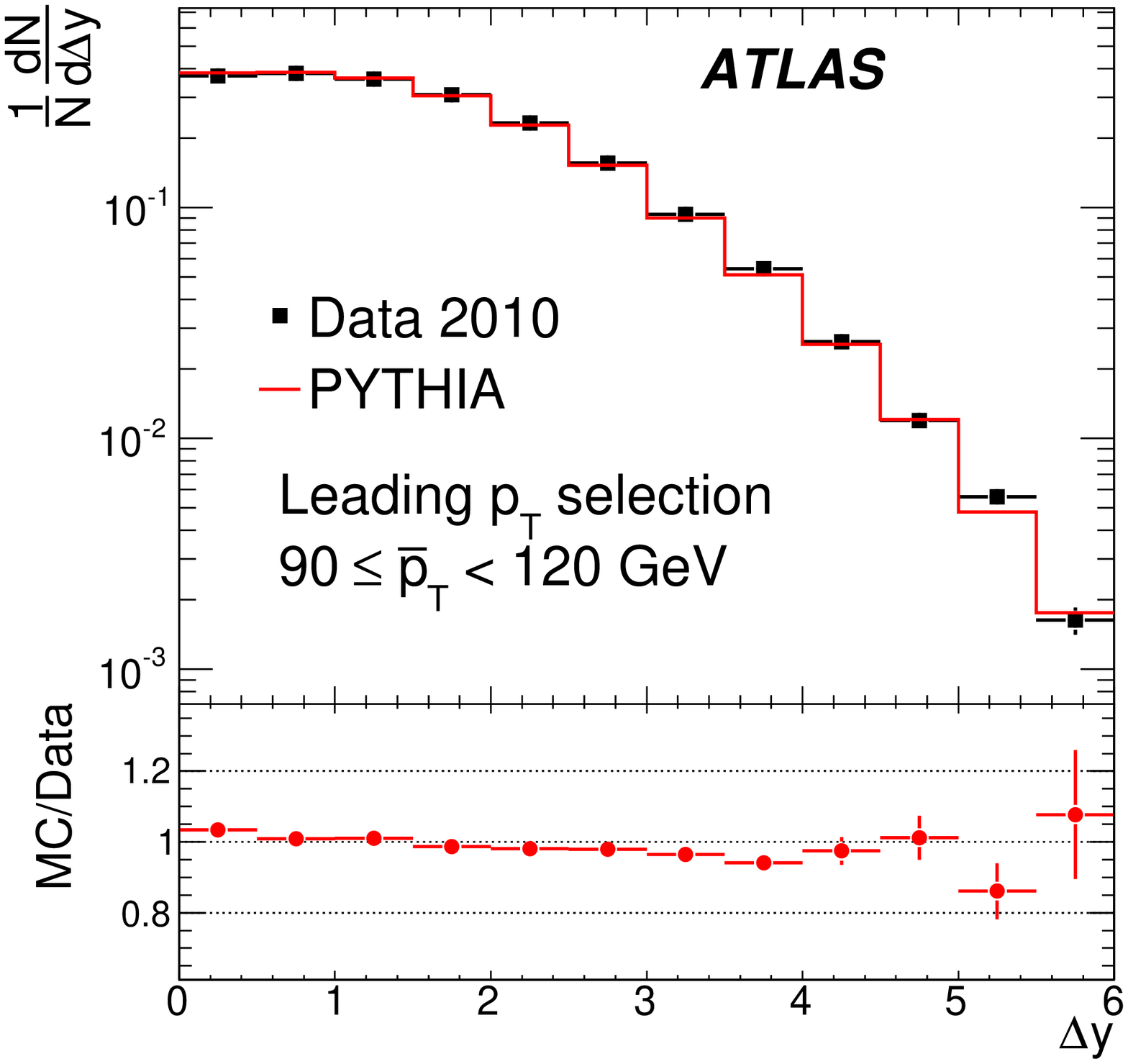}}
  \subfigure[]{ \includegraphics[width=.32\textwidth]{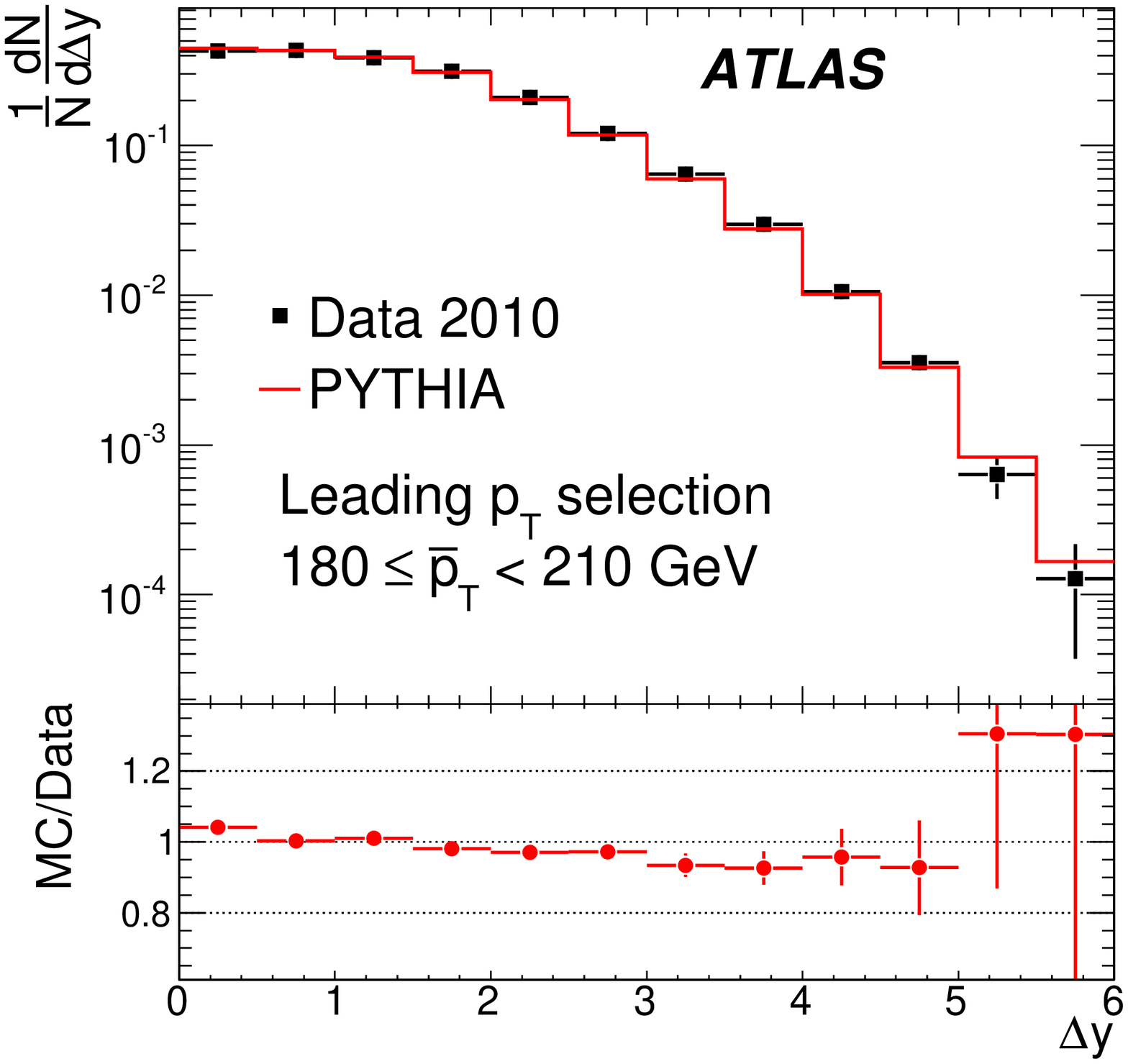}}
  \subfigure[]{ \includegraphics[width=.32\textwidth]{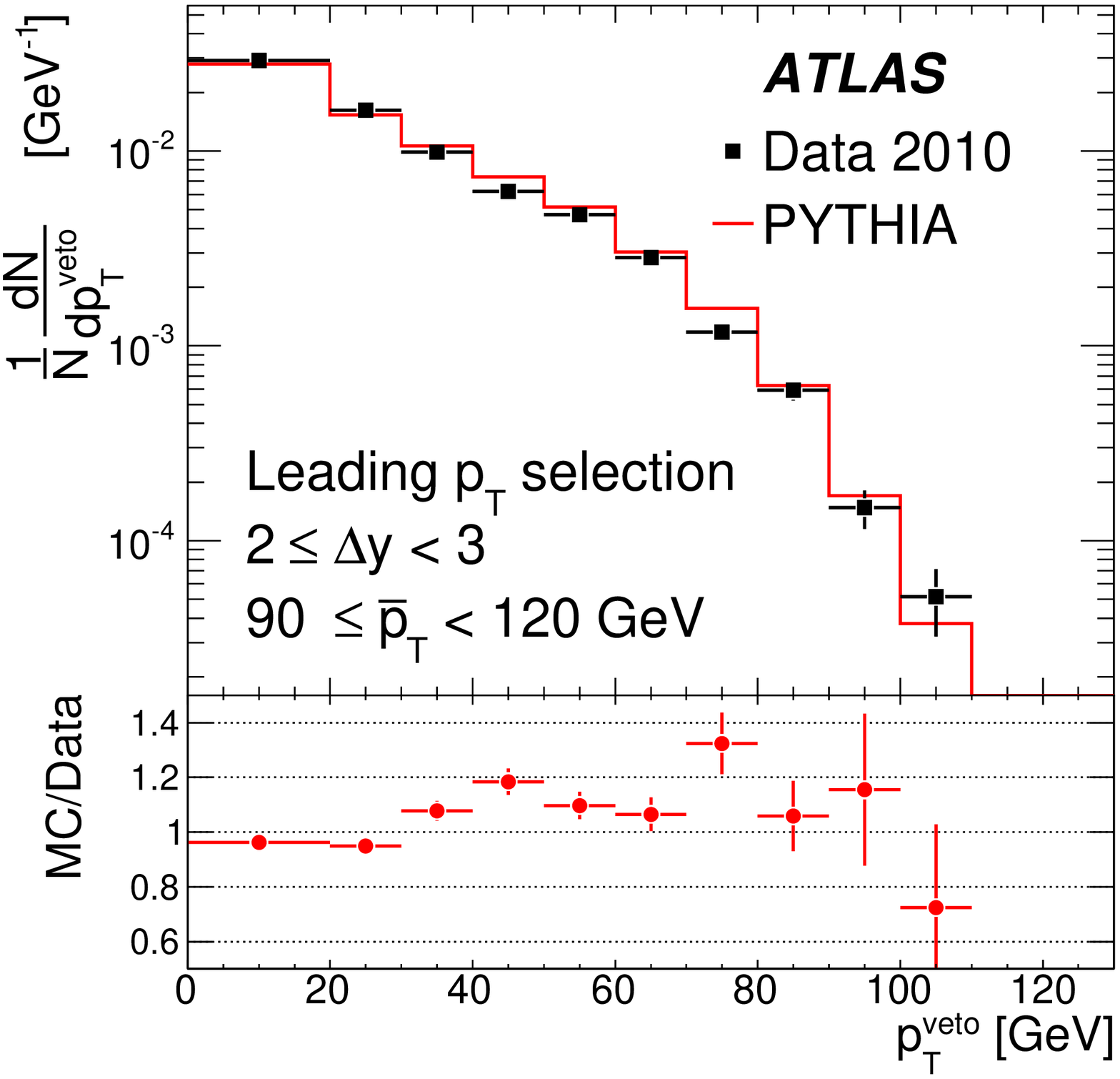}} 
  }
  \mbox{
 \subfigure[]{ \includegraphics[width=.32\textwidth]{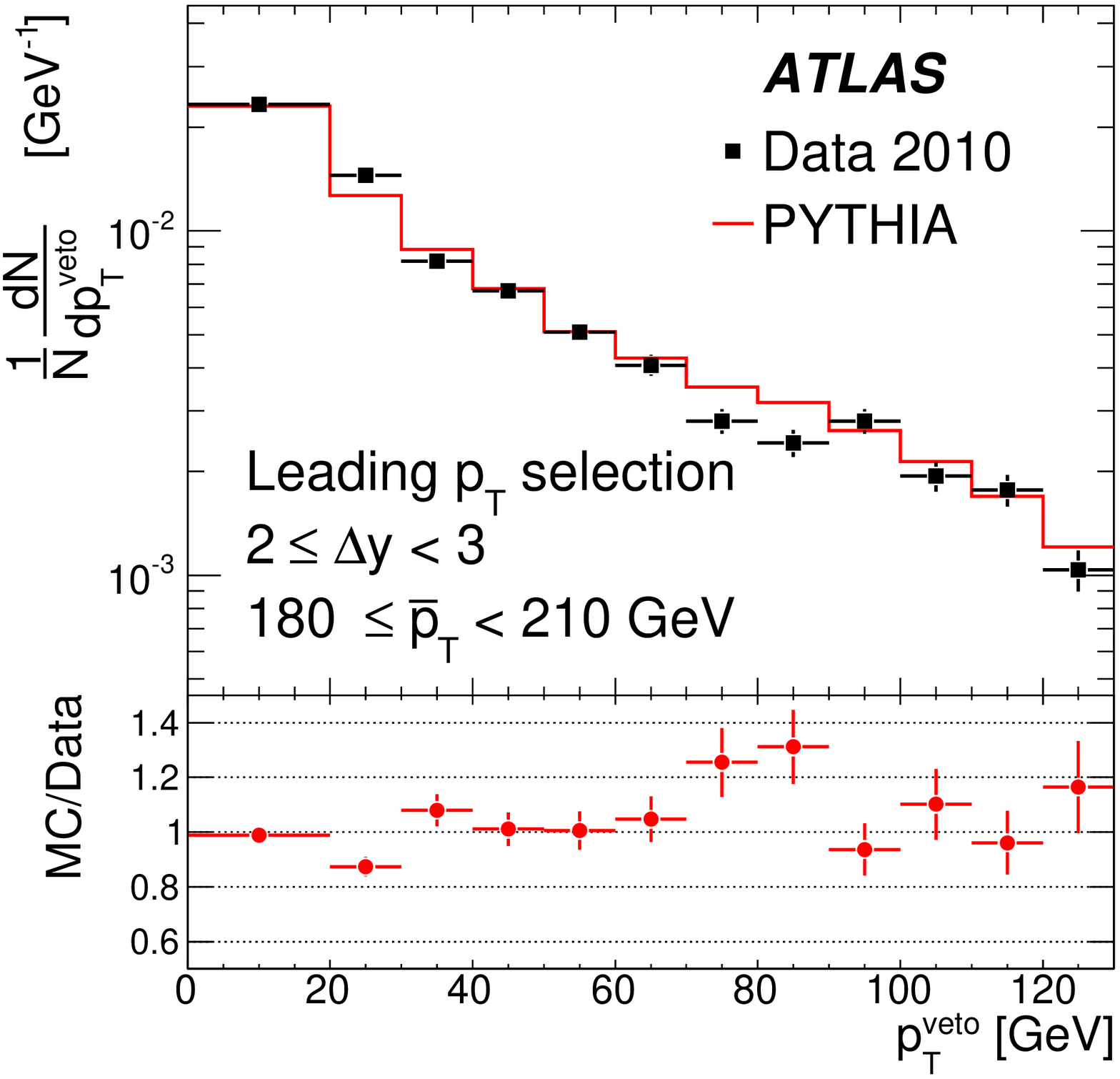}}
 \subfigure[]{ \includegraphics[width=.32\textwidth]{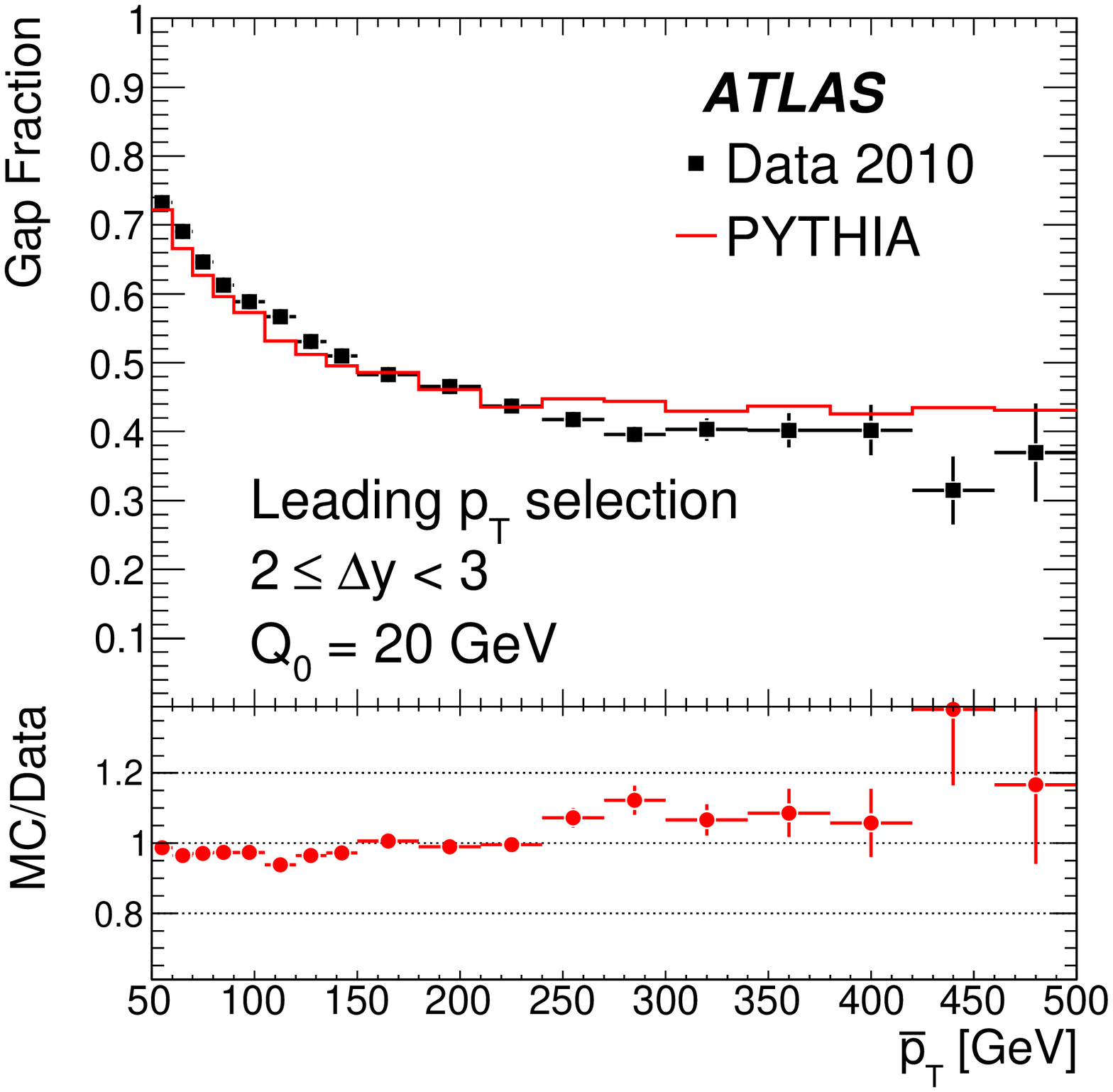}}
 \subfigure[]{ \includegraphics[width=.32\textwidth]{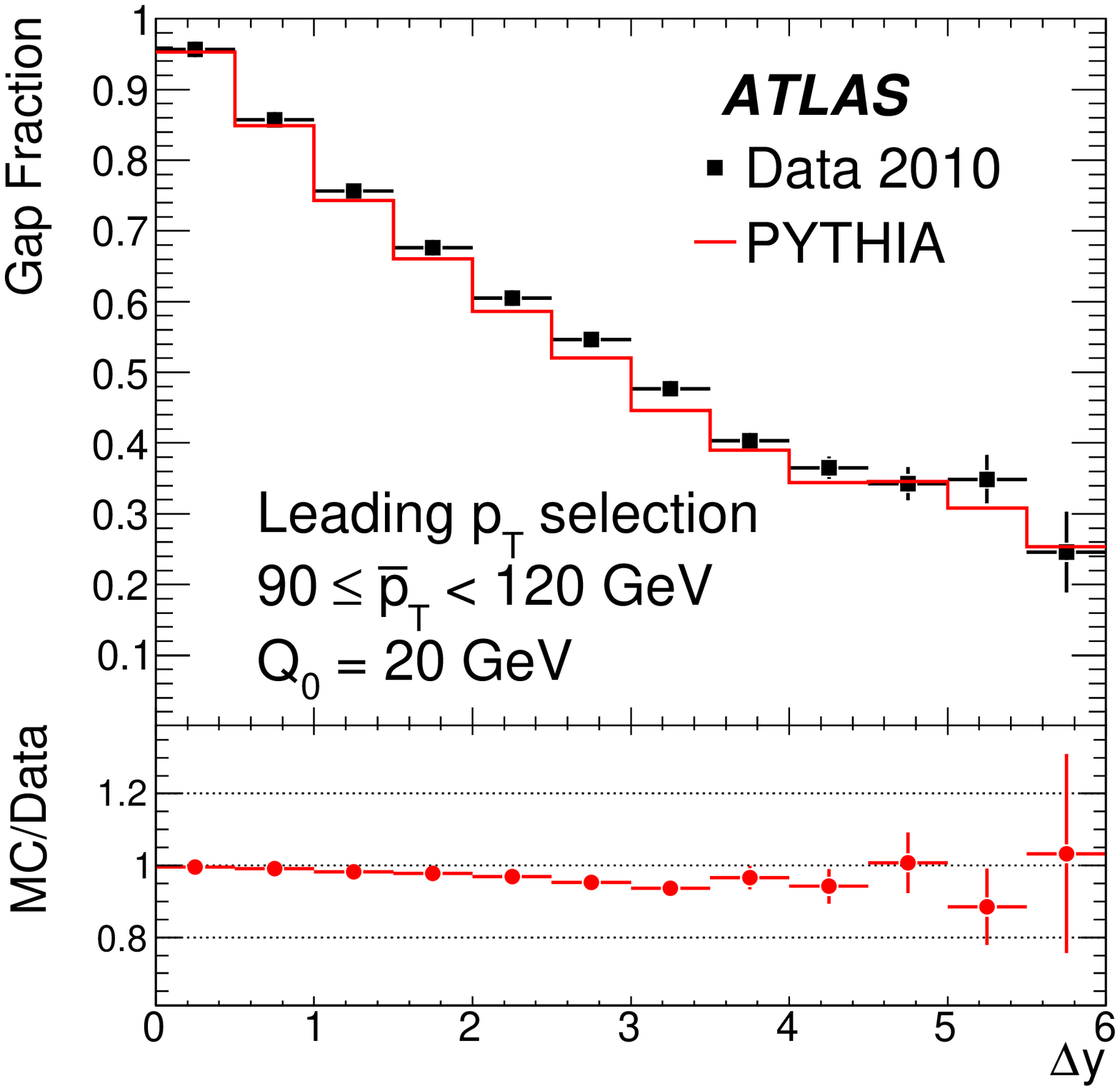}}
 }
  \caption{Control distributions comparing uncorrected data and the \pythia{} MC (tune AMBT1) with \geant{} detector simulation. The dijet system is defined as the two leading-$p_{\rm T}$ jets in the event. The rapidity interval between those jets is shown for the phase space regions  $90 \leq \overline{p}_{\rm T}<120$~GeV and $180 \leq \overline{p}_{\rm T}<210$~GeV in (a) and (b), respectively. The transverse momentum of the leading jet in this rapidity interval, $p_{\rm T}^{\rm veto}$ is shown for $90 \leq \overline{p}_{\rm T}<120$~GeV and $2 \leq \Delta y<3$ in (c) and for $180 \leq \overline{p}_{\rm T}<210$~GeV and $2 \leq \Delta y<3$ in (d). The gap fraction is shown as a function of  $\overline{p}_{\rm T}$ and $\Delta y$ in (e) and (f), respectively. \label{fig:uncorr}}
\end{figure}

In total, 533063 events pass the selection criteria and kinematic cuts if the dijet system is defined as the two leading-$p_{\rm T}$ jets in the event. Of these, 85546 events have $\overline{p}_{\rm T}^{}>$~210~GeV, for which the unprescaled jet triggers are used. If the dijet system is defined as the most forward and most backward jets in the event, then 306364 events pass the selection criteria and kinematic cuts, with 33997 events satisfying $\overline{p}_{\rm T}^{}>$~210~GeV.

The distribution of dijet events as a function of the rapidity interval between the two jets is presented for uncorrected data in Figs.~\ref{fig:uncorr} (a) and (b) for the regions $90 \leq \overline{p}_{\rm T} < 120$~GeV and $180 \leq \overline{p}_{\rm T} < 210$~GeV, respectively. The dijet system is defined as the two leading transverse momentum jets in the event. 
The transverse momentum of the leading jet in the rapidity interval bounded by the dijet system, $p_{\rm T}^{\rm veto}$, is presented in Figs.~\ref{fig:uncorr} (c) and (d). Finally the gap fraction is presented as a function of $\overline{p}_{\rm T}$ and as a function of $\Delta y$ in Figs.~\ref{fig:uncorr} (e) and (f), respectively. In all such distributions, the baseline \pythia{} event generator with \geant{} detector simulation gives a reasonable description of the uncorrected data.

\section{Correction for detector effects}\label{sec:corr}
The corrections for detector effects were calculated with a bin-by-bin unfolding procedure. In this approach, the correction is defined in each bin as the ratio of the hadron-level distribution (including muons and neutrinos) to the detector level distributions, using the \pythia{} event generator. The bin sizes were chosen to be commensurate with the jet energy resolution to ensure that the bin-to-bin migration was not too large. In particular, the purity\footnote{The bin purity is calculated using \pythia{} events and defined as the number of events that are both reconstructed and generated in a particular bin divided by the number of events that are reconstructed in that bin.} of each bin was required to be at least 50\% (the typical bin purity was between 60\% and 70\%). 
The typical correction factor was observed to be a few percent for the gap fraction distribution and between 5\% and 10\% for the distribution of mean number of jets in the rapidity interval between the boundary jets. 

The systematic uncertainty on the detector correction was determined in two steps. The physics modelling uncertainty was estimated by reweighting the $\overline{p}_{\rm T}$, $\Delta y$ and $p_{\rm T}^{\rm veto}$ distributions to account for any deviation between data and MC and also to cover the maximal variation in shape allowed by the JES uncertainty. 
The detector modelling uncertainty was determined by reweighting the $z$-vertex distribution and varying the jet reconstruction efficiency and the jet energy resolution within the allowed uncertainties as determined from data \cite{reseff}. The modelling uncertainties were cross-checked using the \herwigpp{} and \alpgen{} samples, which agreed with the baseline \pythia{} values within the statistical uncertainty of each sample. The total systematic uncertainty in the detector correction was defined as the quadrature sum of the physics/detector modelling uncertainties and the statistical uncertainty of the \pythia{} samples. The systematic uncertainty on the correction procedure is typically 2-3\%. This uncertainty increases when $\Delta y$ and $\overline{p}_{\rm T}$ are both large, due to an increased statistical uncertainty in the MC samples, and the maximum uncertainty is about 10\% at $\Delta y \sim 5$ and $\overline{p}_{\rm T} \sim 240$~GeV. This does not have a detrimental impact on the measurement, however, as the statistical uncertainty on the data in these regions of phase space is much larger.
\section{Results and discussion}\label{sec:gfres}
\begin{figure}[t]
\centering
\mbox{
  \subfigure[]{ \includegraphics[width=.48\textwidth]{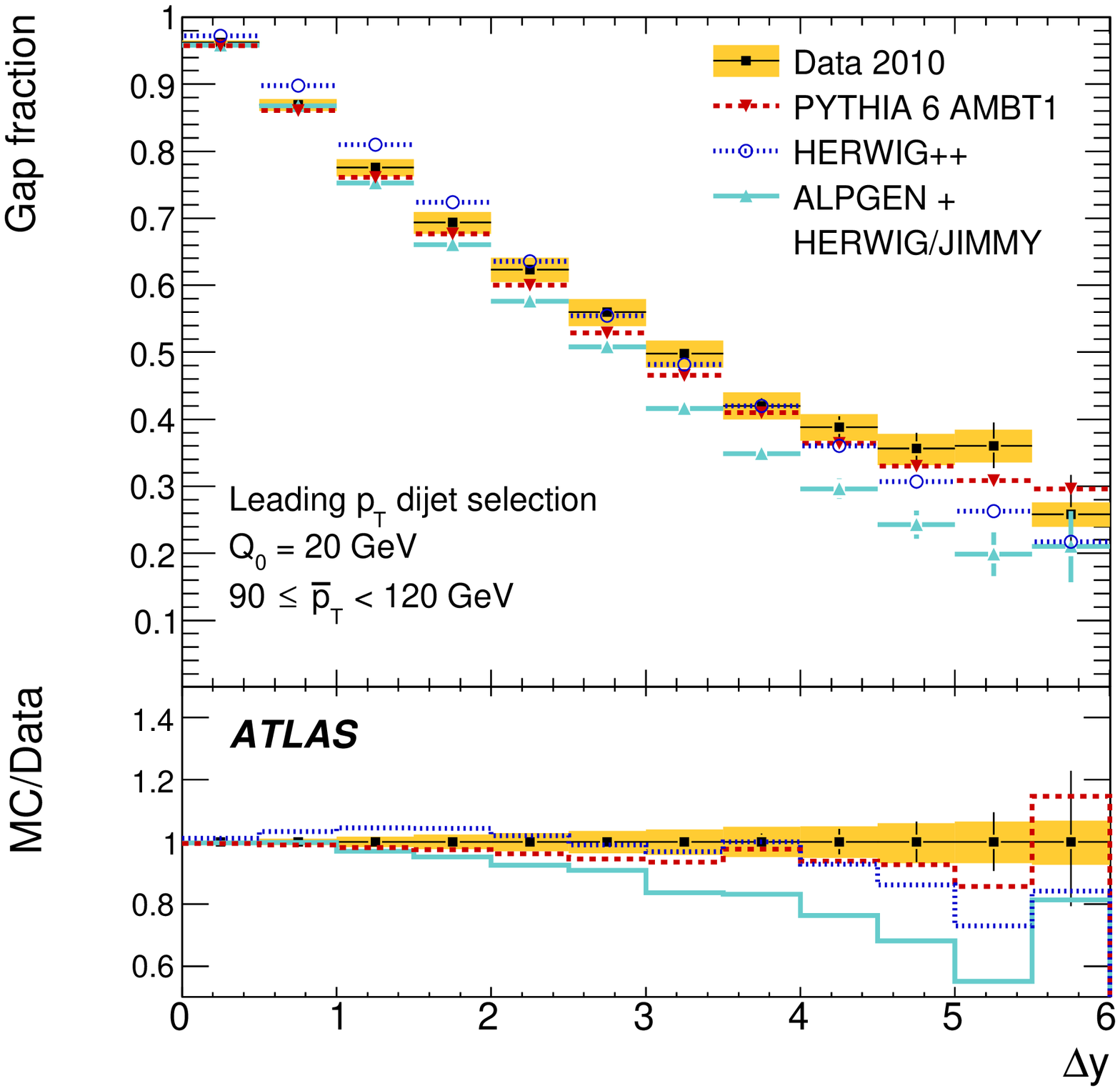}}
  \subfigure[]{ \includegraphics[width=.48\textwidth]{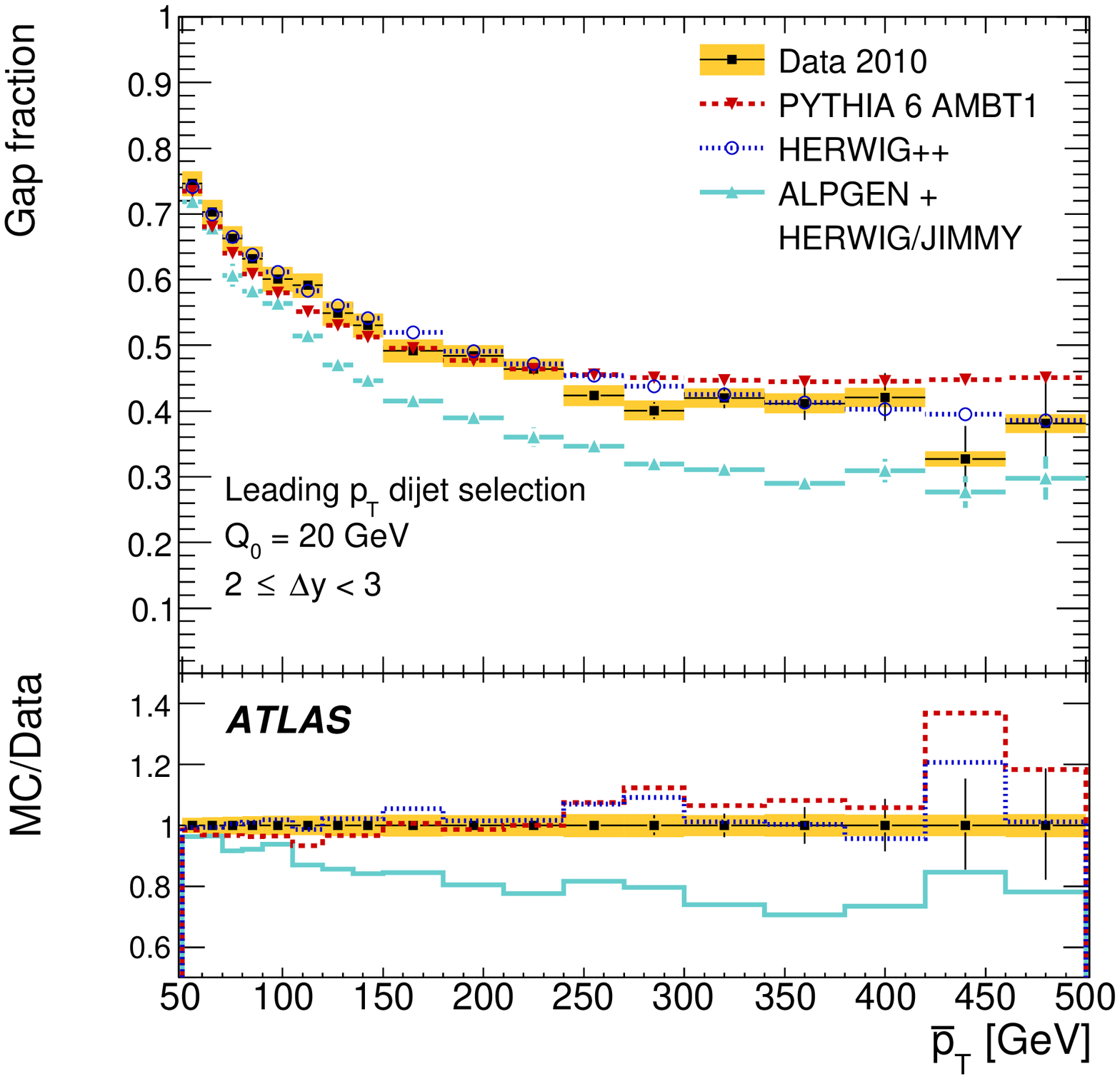}}
 }
 \caption{ Gap fraction as a function of $\Delta y$, given that the dijet system is defined as the leading-$p_{\rm T}^{}$ jets in the event and satisfies $90 \leq \overline{p}_{\rm T}<120$~GeV (a). Gap fraction as a function of $\overline{p}_{\rm T}$ given that the rapidity interval is $2 \leq \Delta y<3$ (b). The (corrected) data are the black points, with error bars representing the statistical uncertainty. The total systematic uncertainty on the measurement is represented by the solid (yellow) band. The dashed (red) points represents the \pythia{} prediction (tune AMBT1), the dot-dashed (blue) points represents the \herwigpp{} prediction (tune LHC-UE7-1) and the solid (cyan) points represents the \alpgen{} prediction (tune AUET1).  \label{fig:mc_gf}}
\end{figure}

Figure \ref{fig:mc_gf} shows the gap fraction as a function of $\Delta y$ and $\overline{p}_{\rm T}$, with the data compared to the \pythia{}, \herwigpp{} and \alpgen{} event generators. The dijet system is defined as the two leading-$p_{\rm T}$ jets in the event. The data are corrected for detector effects, as discussed in Sec.~\ref{sec:corr}. The total uncertainty due to systematic effects is the sum in quadrature of the uncertainty due to JES (Sec.~\ref{sec:jes}) and the uncertainty due to the correction for detector effects  (Sec.~\ref{sec:corr}). All other systematic effects were determined to have negligible impact and were therefore not included in the final systematic uncertainty. Both \pythia{} and \herwigpp{} give a good description of the data as a function of $\overline{p}_{\rm T}$. \pythia{} also gives the best description of the data as a function of $\Delta y$, although the gap fraction is slightly underestimated for $\Delta y \sim 3$. \herwigpp{} overestimates the gap fraction at low values of $\Delta y$ and underestimates the gap fraction at large values of $\Delta y$. \alpgen{} shows the largest deviation from the data, predicting a gap fraction that is too small at large values of $\Delta y$ or $\overline{p}_{\rm T}$. 

\begin{figure}[t]
\centering
\mbox{
    \subfigure[]{ \includegraphics[width=.5\textwidth]{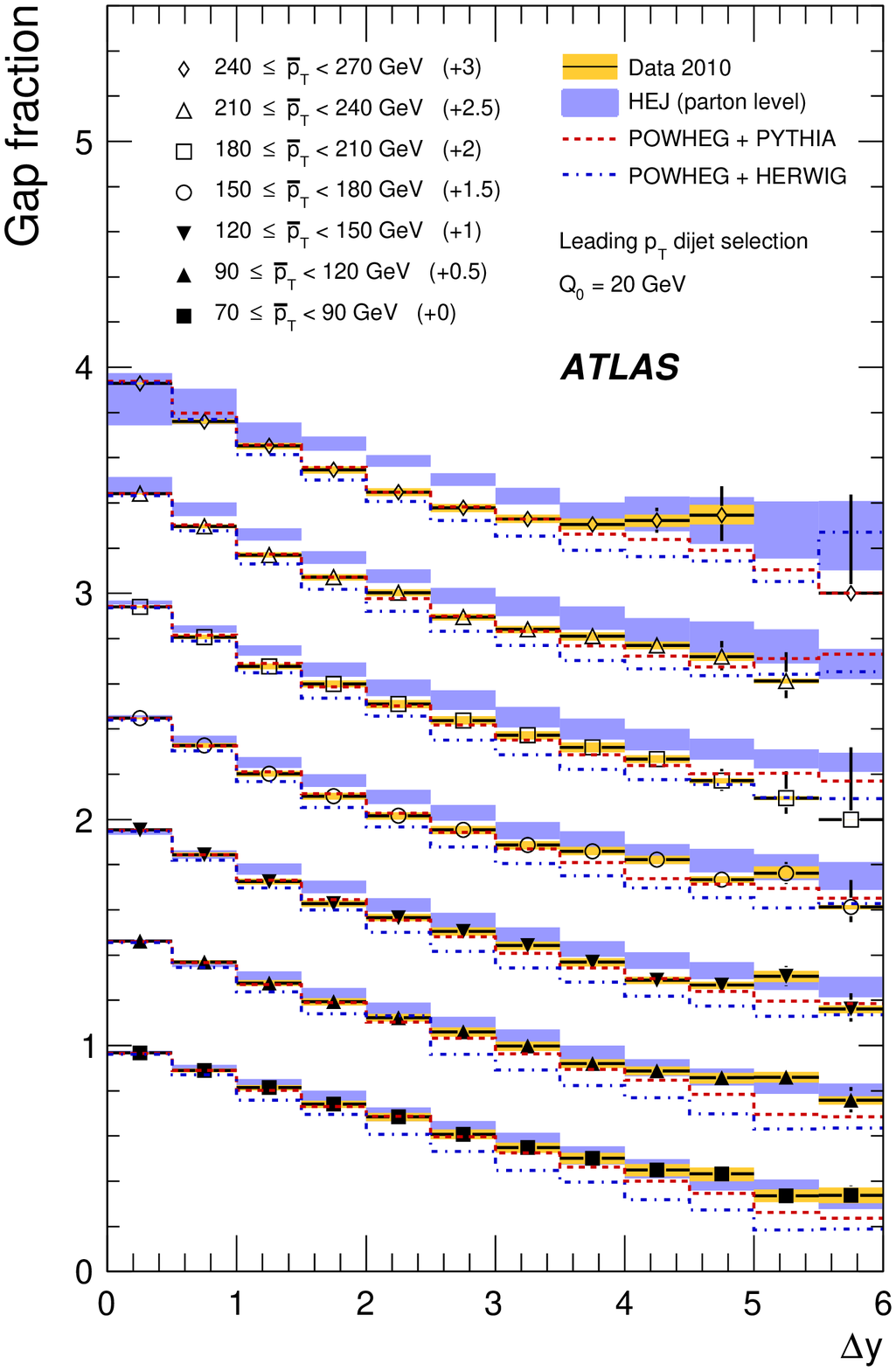}}
  \subfigure[]{ \includegraphics[width=.5\textwidth]{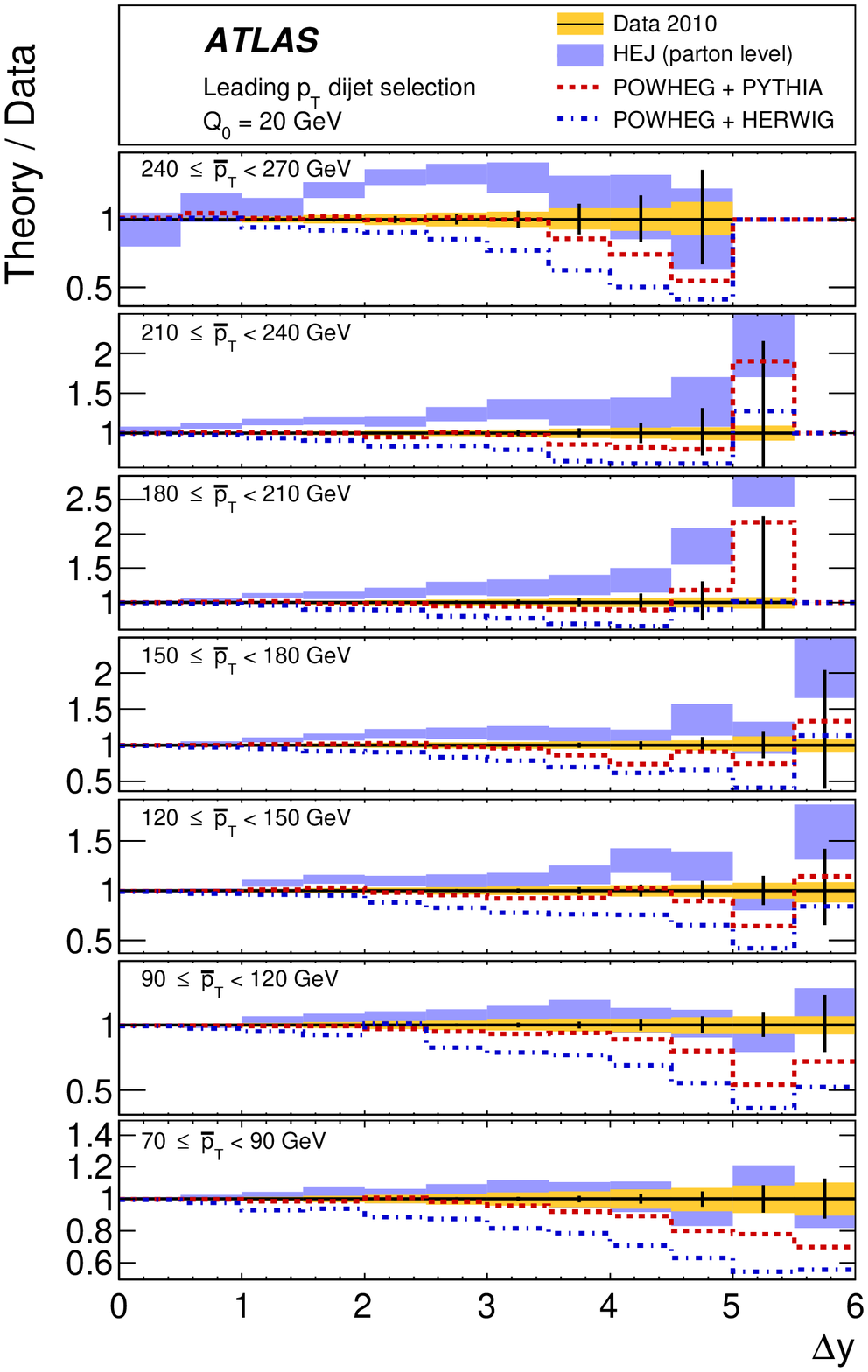}}
  }  
\caption{Gap fraction as a function of $\Delta y$ for various $\overline{p}_{\rm T}$ slices. The dijet system is defined as the two leading-$p_{\rm T}^{}$ jets in the event. The data are compared to the \hej{} and \powheg{} predictions in (a). The ratio of these theory predictions to the data are shown in (b). The (unfolded) data are the black points, with error bars representing the statistical uncertainty and a solid (yellow) band representing the total systematic uncertainty. The darker (blue) band represents the theoretical uncertainty in the \hej{} calculation from variation of the PDF and renormalisation/factorisation scales. 
The dashed (red) and dot-dashed (blue) curves represent the \powheg{} predictions after showering, hadronisation and underlying event simulation with \pythia{} (tune AMBT1) and \herwig{}/\jimmy{} (tune AUET1), respectively.\label{fig:gf_vs_ptbarA}}
\end{figure}

\begin{figure}[t]
\centering
\mbox{
  \subfigure[]{\includegraphics[width=0.5\textwidth]{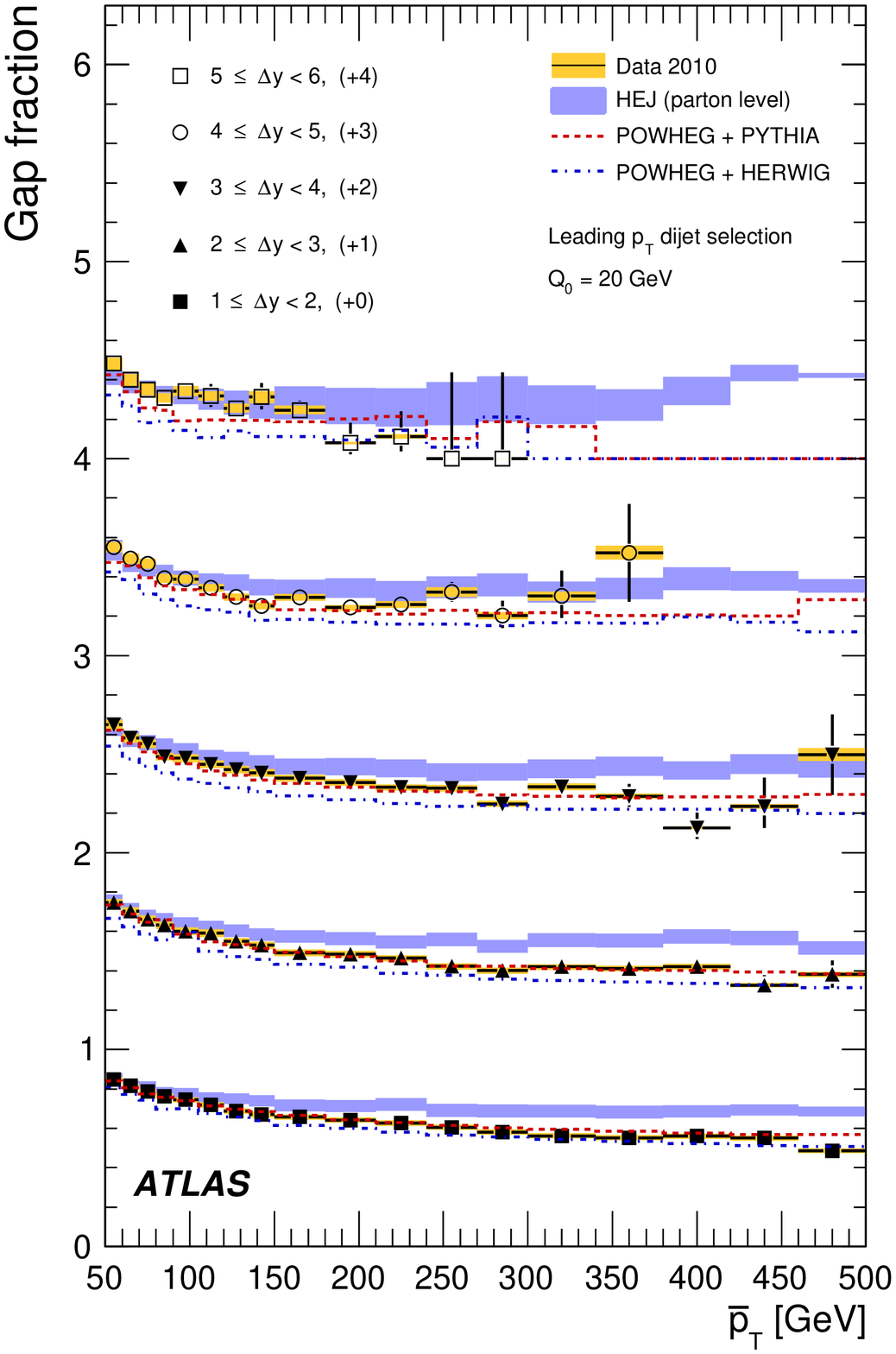}}
  \subfigure[]{\includegraphics[width=0.5\textwidth]{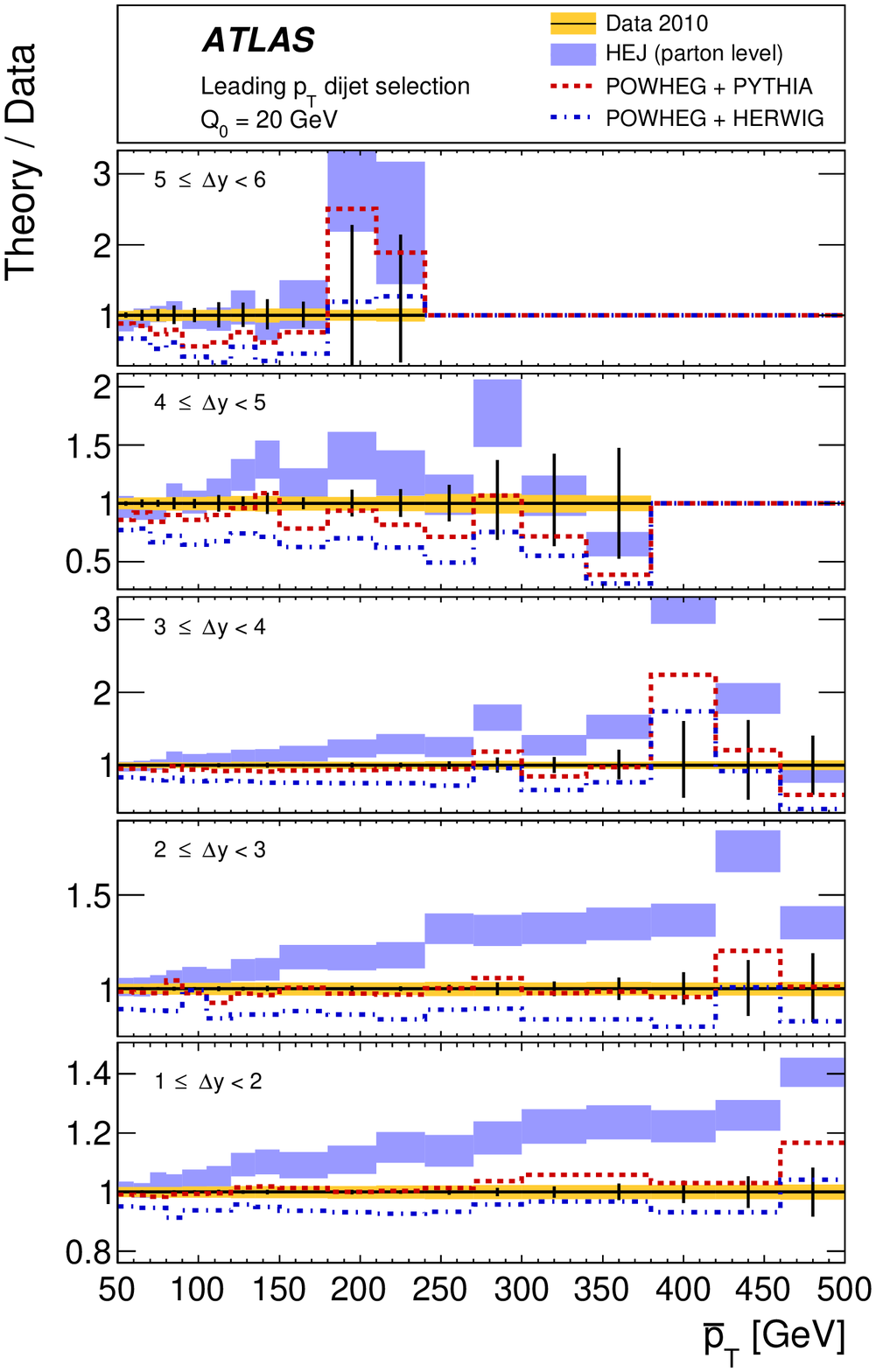}}
  }
\caption{Gap fraction as a function of  $\overline{p}_{\rm T}$ for various $\Delta y$  slices. The dijet system is defined as the two leading-$p_{\rm T}^{}$ jets in the event. The data are compared to the \hej{} and \powheg{} predictions in (a). The ratio of these theory predictions to the data are shown in (b). 
The data and theory are presented in the same way as Fig.~\ref{fig:gf_vs_ptbarA}. 
\label{fig:gf_vs_dyA}}
\end{figure}

The data are compared to the \hej{} and \powheg{} predictions in Fig.~\ref{fig:gf_vs_ptbarA} and Fig.~\ref{fig:gf_vs_dyA} as a function of $\Delta y$ and $\overline{p}_{\rm T}$, respectively. The dijet system is again defined as the two leading-$p_{\rm T}$ jets in the event.  The dependence of the gap fraction on one variable is studied after fixing the phase space of the other variable to well defined and narrow regions. The \hej{} prediction describes the data well as a function of $\Delta y$ at low values of $\overline{p}_{\rm T}$. However, at large values of $\overline{p}_{\rm T}$, \hej{} predicts too many gap events. It should be noted that \hej{} is designed to give a good description of QCD in the limit that all the jets have similar transverse momentum. Therefore, the failure of the \hej{} calculation as $\overline{p}_{\rm T}$ becomes much larger than $Q_{0}$ is not unexpected. The description of the data may be improved by matching the \hej{} calculation to a standard parton shower, to account for soft and collinear emissions \cite{HEJshowered}. 

In general, \powhegpyt{} provides the best description of the data, when considered over all the phase space regions presented. However, at large values of $\Delta y$, the gap fraction predicted by \powhegpyt{} deviates from the data. This is expected because the NLO-plus-parton-shower approximation does not contain the contributions to a full QCD calculation that become important as  $\Delta y$ increases. The gap fraction as a function of $\overline{p}_{\rm T}$ is, however, well described by \powhegpyt{} at low $\Delta y$. 
Furthermore, although the absolute value of the gap fraction is not correct at larger $\Delta y$, the shape of the distributions in $\overline{p}_{\rm T}$ remain well described. \powhegher{} tends to produce too much activity across the full phase-space. However, the difference between \powhegher{} and the data increases with $\Delta y$, reproducing the effect observed with \powhegpyt{}. 

\begin{figure}[t]
\centering
\mbox{
  \subfigure[]{ \includegraphics[width=.5\textwidth]{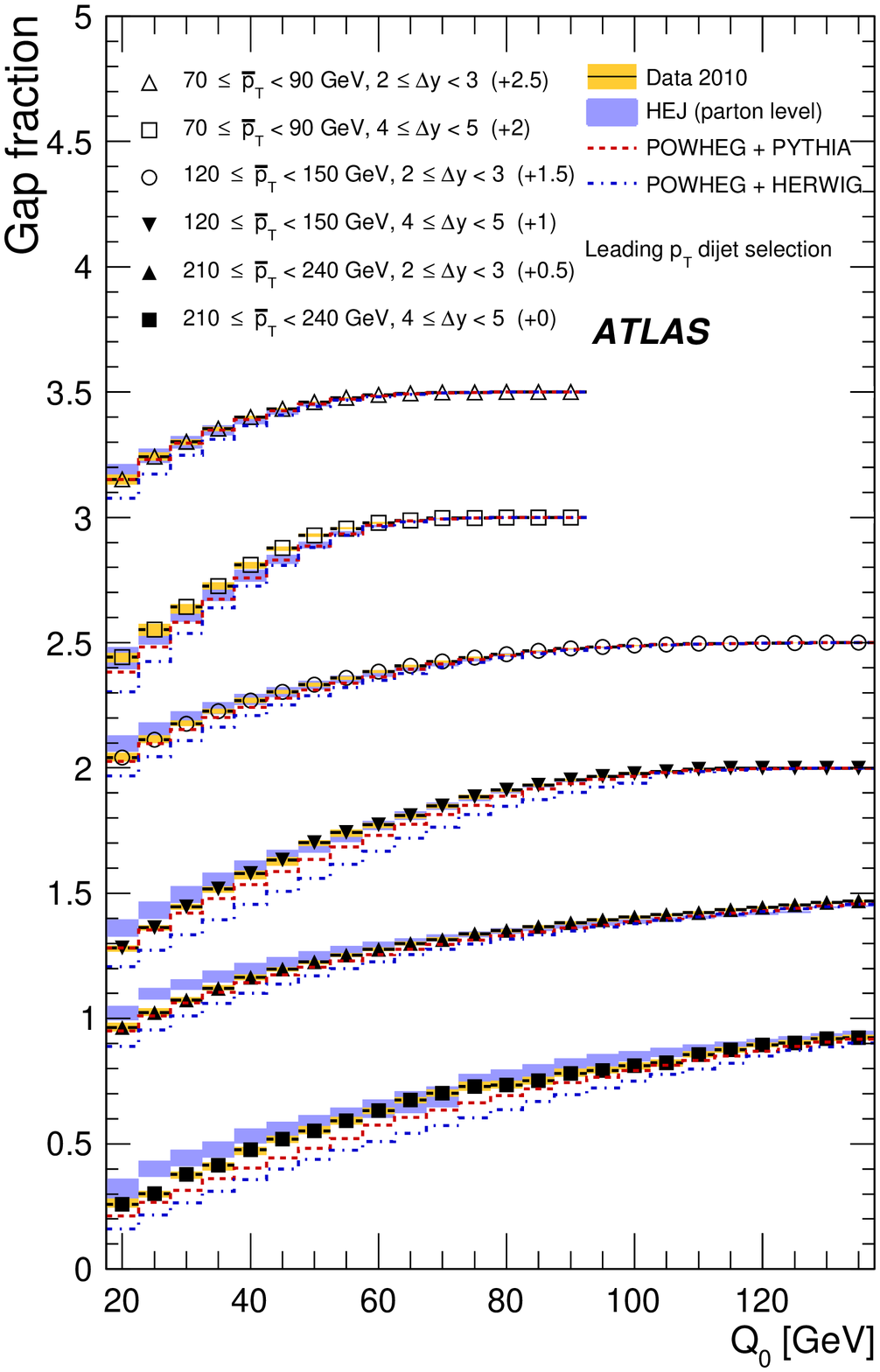}}
  \subfigure[]{ \includegraphics[width=.5\textwidth]{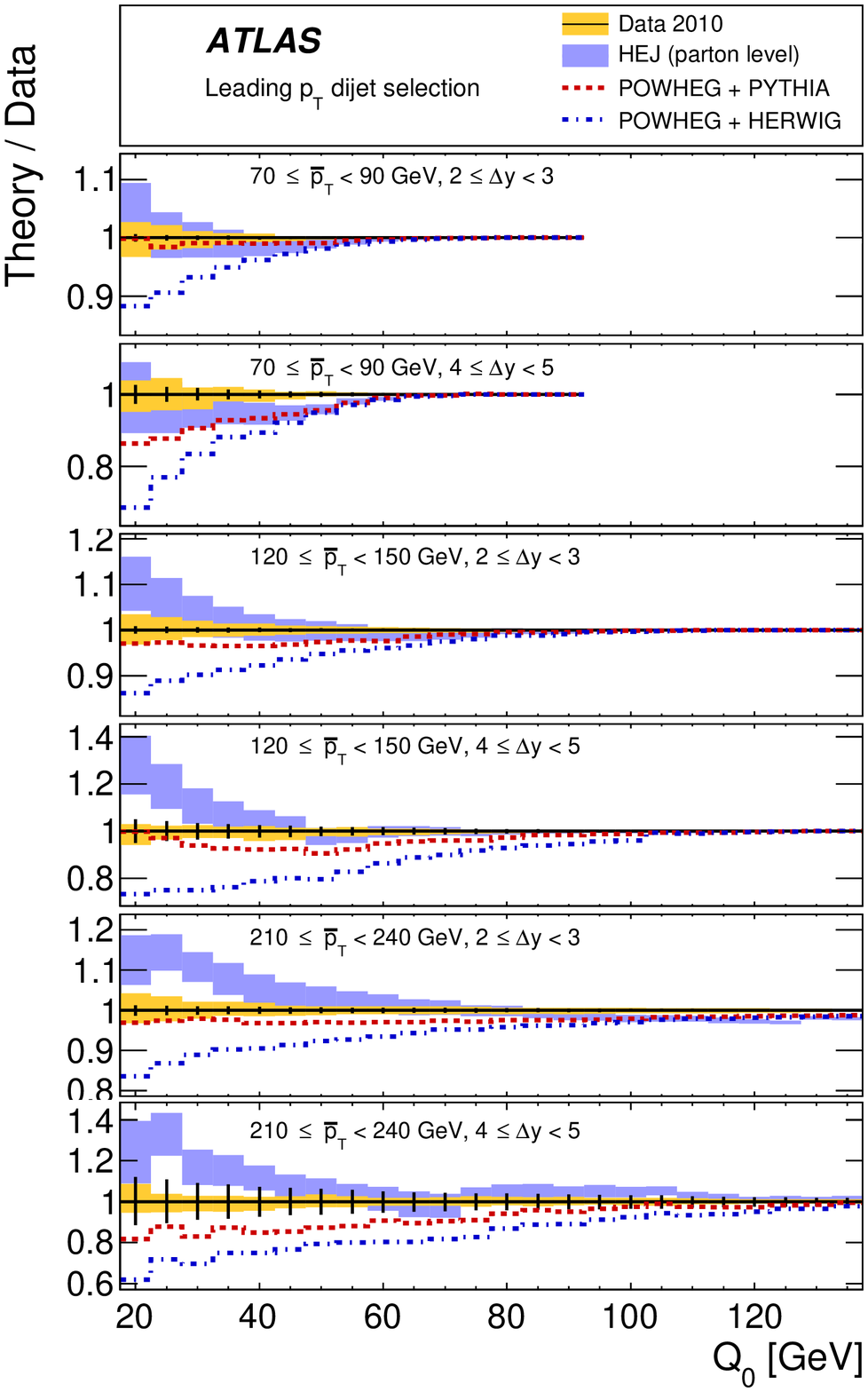}}
  }  
\caption{Gap fraction as a function of  $Q_{0}$ for various $\overline{p}_{\rm T}$ and $\Delta y$ slices. 
The dijet system is defined as the two leading-$p_{\rm T}^{}$ jets in the event. The data are compared to the \hej{} and \powheg{} predictions in (a). The data points for $Q_{0}>\overline{p}_{\rm T}$ have been removed because the gap fraction is always equal to one for this dijet selection, by definition. The ratio of the theory predictions to the data are shown in (b). The data and theory are presented in the same way as Fig.~\ref{fig:gf_vs_ptbarA}.\label{fig:gf_vs_q0a}}
\end{figure}

The dependence of the gap fraction on the veto scale is presented in Fig.~\ref{fig:gf_vs_q0a} for specific regions of $\overline{p}_{\rm T}$ and $\Delta y$. The $Q_{0}$ dependence of the cross-section is useful in studying the colour structure of the event \cite{Schofield:2011zi}. The difference between \powhegpyt{} and \powhegher{} remains large for all values of $Q_{0}$. The \hej{} description of the data  improves as $Q_{0}$ approaches $\overline{p}_{\rm T}$, a kinematic configuration more suited to the \hej{} approximations. At large values of $\overline{p}_{\rm T}$, none of the theoretical predictions describe the data well as a function of $Q_{0}$. In particular, the description of the data is particularly poor when both $\overline{p}_{\rm T}$ and $\Delta y$ become large, corresponding to the region in which colour singlet exchange is expected to play an increasingly important role.

\begin{figure}[t]
\centering
\mbox{
  \subfigure[]{ \includegraphics[width=.5\textwidth]{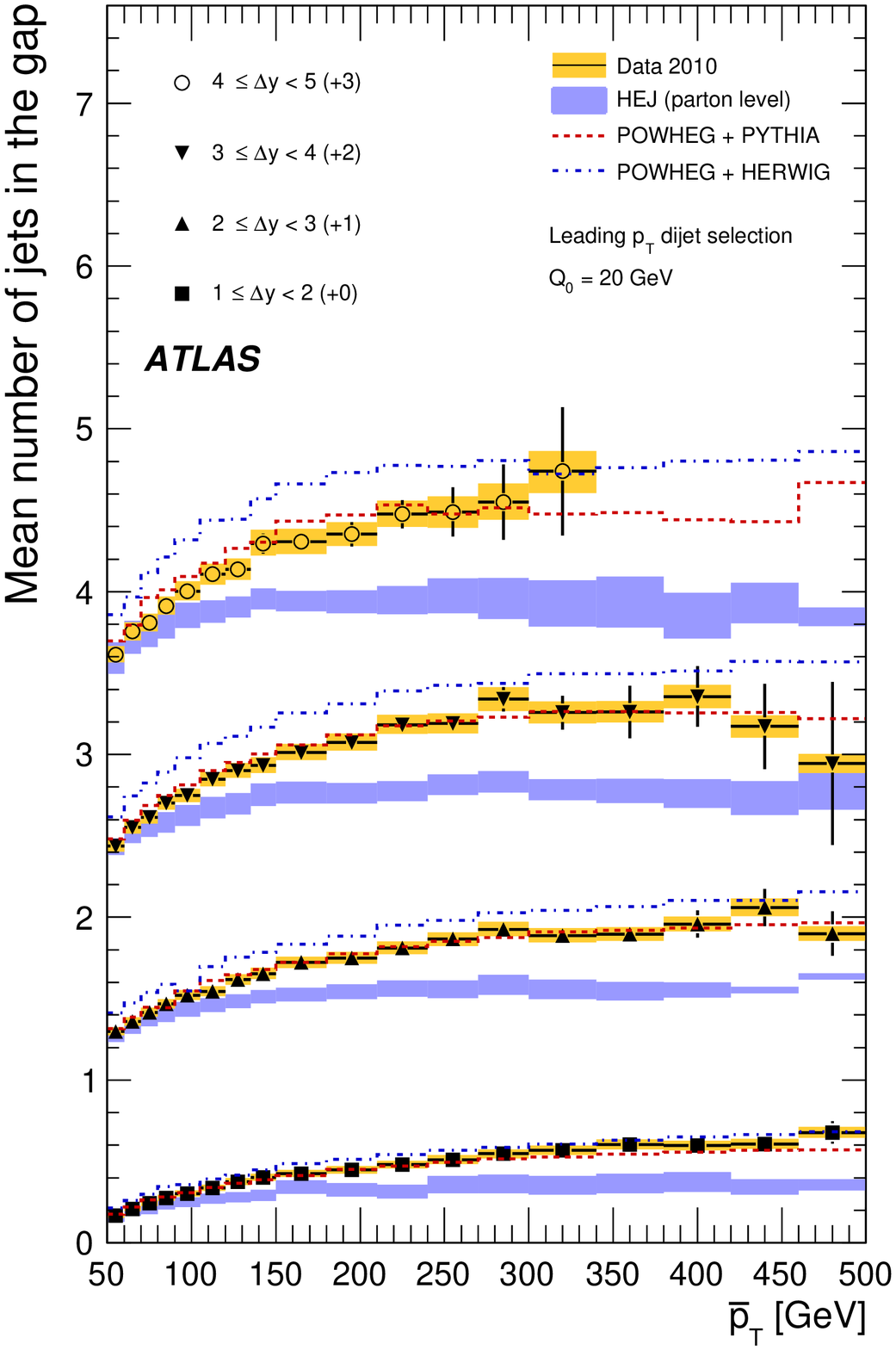}}
  \subfigure[]{ \includegraphics[width=.5\textwidth]{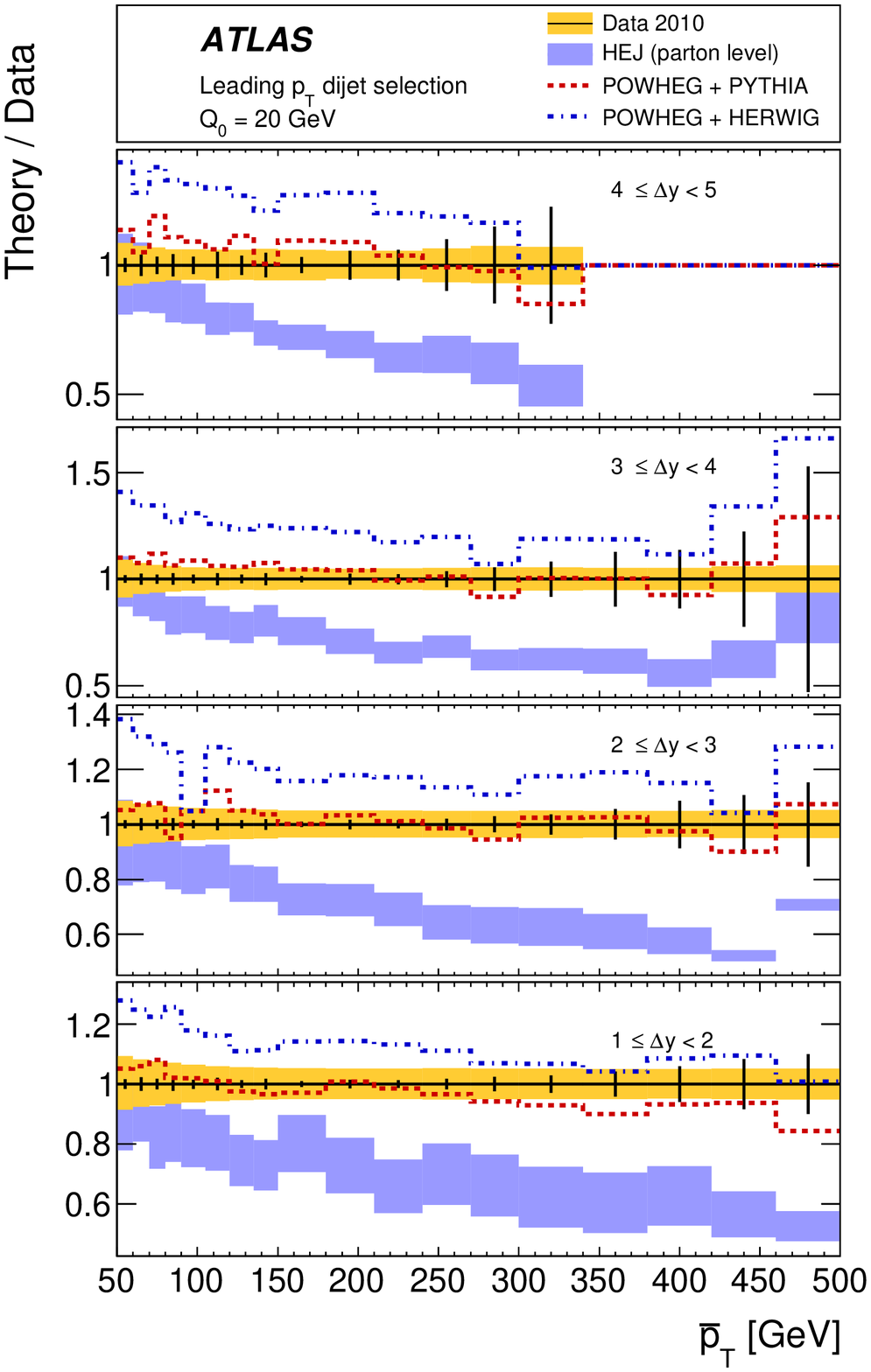}}
  }  
\caption{Mean number of jets in the gap as a function of $\overline{p}_{\rm T}$  for various $\Delta y$ slices. The dijet system is defined as the two leading-$p_{\rm T}^{}$ jets in the event. The data are compared to the \hej{} and \powheg{} predictions in (a). The ratio of these theory predictions to the data are shown in (b).   The data and theory are presented in the same way as Fig.~\ref{fig:gf_vs_ptbarA}.\label{fig:n_vs_ptbarA}}
\end{figure}

Figure \ref{fig:n_vs_ptbarA} shows the mean number of jets in the rapidity interval bounded by the dijet system as a function of $\overline{p}_{\rm T}$. This is an alternative way of studying the activity between the boundary jets. The prediction of \powhegpyt{} again gives the best description of the data, replicating the result obtained using the gap fraction. The \powhegher{} description of the data as a function of $\overline{p}_{\rm T}$ becomes worse as $\overline{p}_{\rm T}$ decreases, which was not observed in the gap fraction distribution. In particular, \powhegher{} predicts a mean jet multiplicity that is too large. The \hej{} prediction deviates from the data at large values of $\overline{p}_{\rm T}$, producing too little jet activity. This is the same effect as was observed using the gap fraction, although the deviations from the data are larger. 

\begin{figure}[t]
\centering
\mbox{
  \subfigure[]{ \includegraphics[width=.5\textwidth]{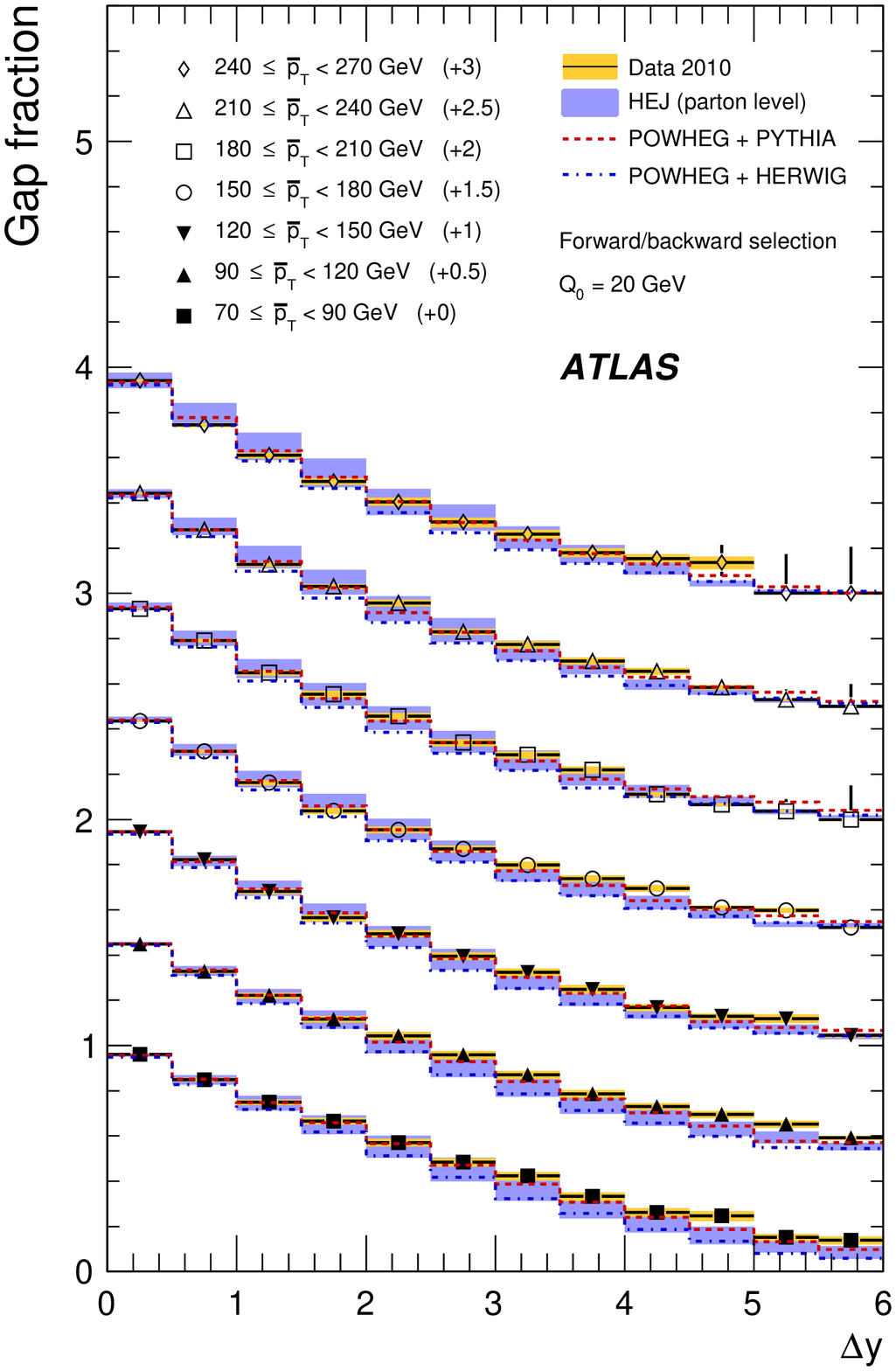}}
  \subfigure[]{ \includegraphics[width=.5\textwidth]{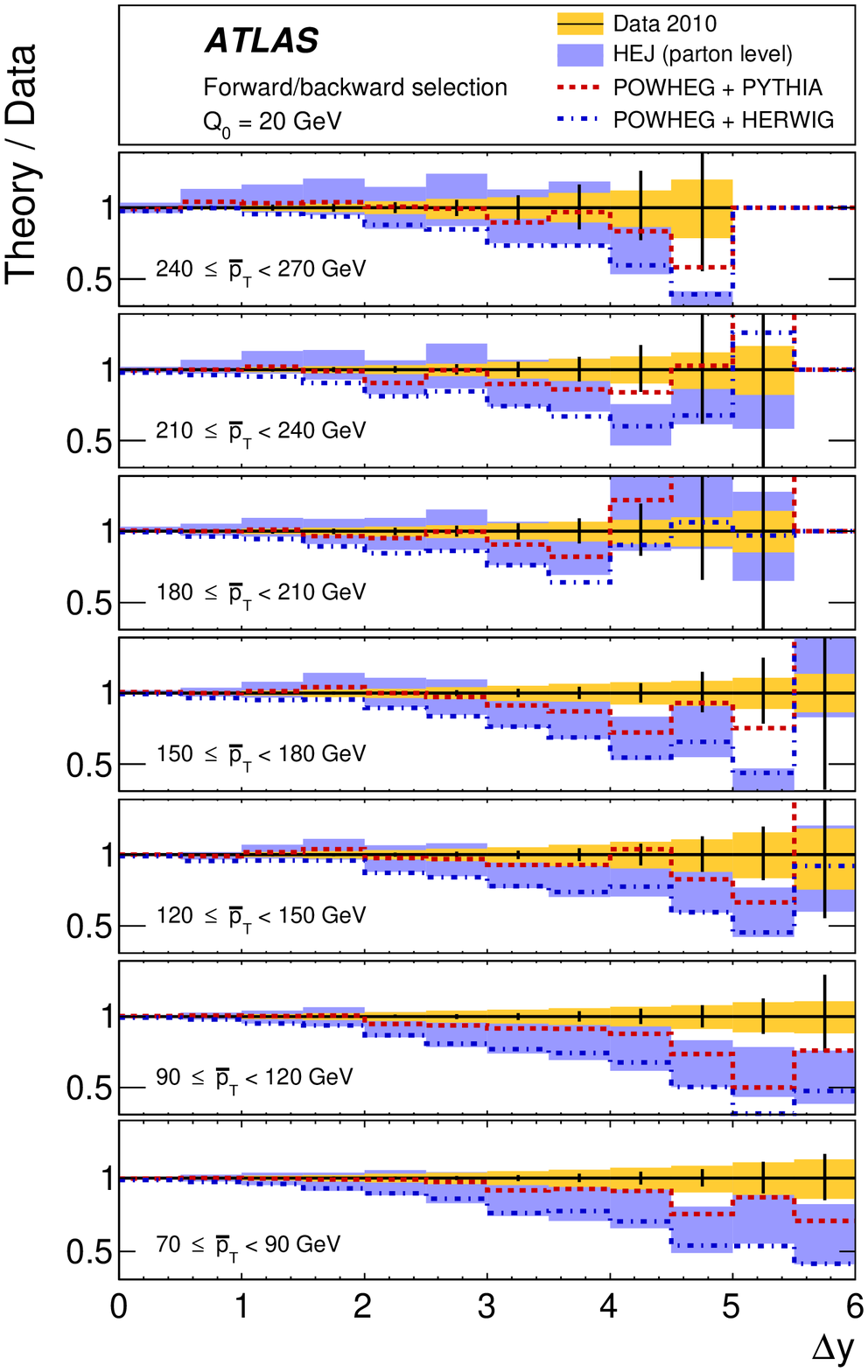}}
  }  
\caption{Gap fraction as a function of  $\Delta y$ for various $\overline{p}_{\rm T}$ slices. The dijet system is defined as the most forward and the most backward jets in the event. The data are compared to the \hej{} and \powheg{} predictions in (a). The ratio of these theory predictions to the data are shown in (b).  The data and theory are presented in the same way as Fig.~\ref{fig:gf_vs_ptbarA}. \label{fig:gf_vs_dyB}}
\end{figure}

Figure \ref{fig:gf_vs_dyB} shows the gap fraction as a function of $\Delta y$, with the dijet system defined as the most forward and the most backward jets in the event. For this selection, the $p_{\rm T}$-imbalance between the two jets is typically much larger than when the dijet system is defined as the two leading-$p_{\rm T}^{}$ jets in the event. The data are not well described by \hej{} at low values of $\overline{p}_{\rm T}$, implying that the resummation of soft emissions are important for this configuration.  The \powheg{} prediction is similar to the \hej{} prediction in all regions of phase space, that is, both calculations result in a gap fraction that is too small at large $\Delta y$.

\begin{figure}[t]
\centering
\mbox{
  \subfigure[]{ \includegraphics[width=.5\textwidth]{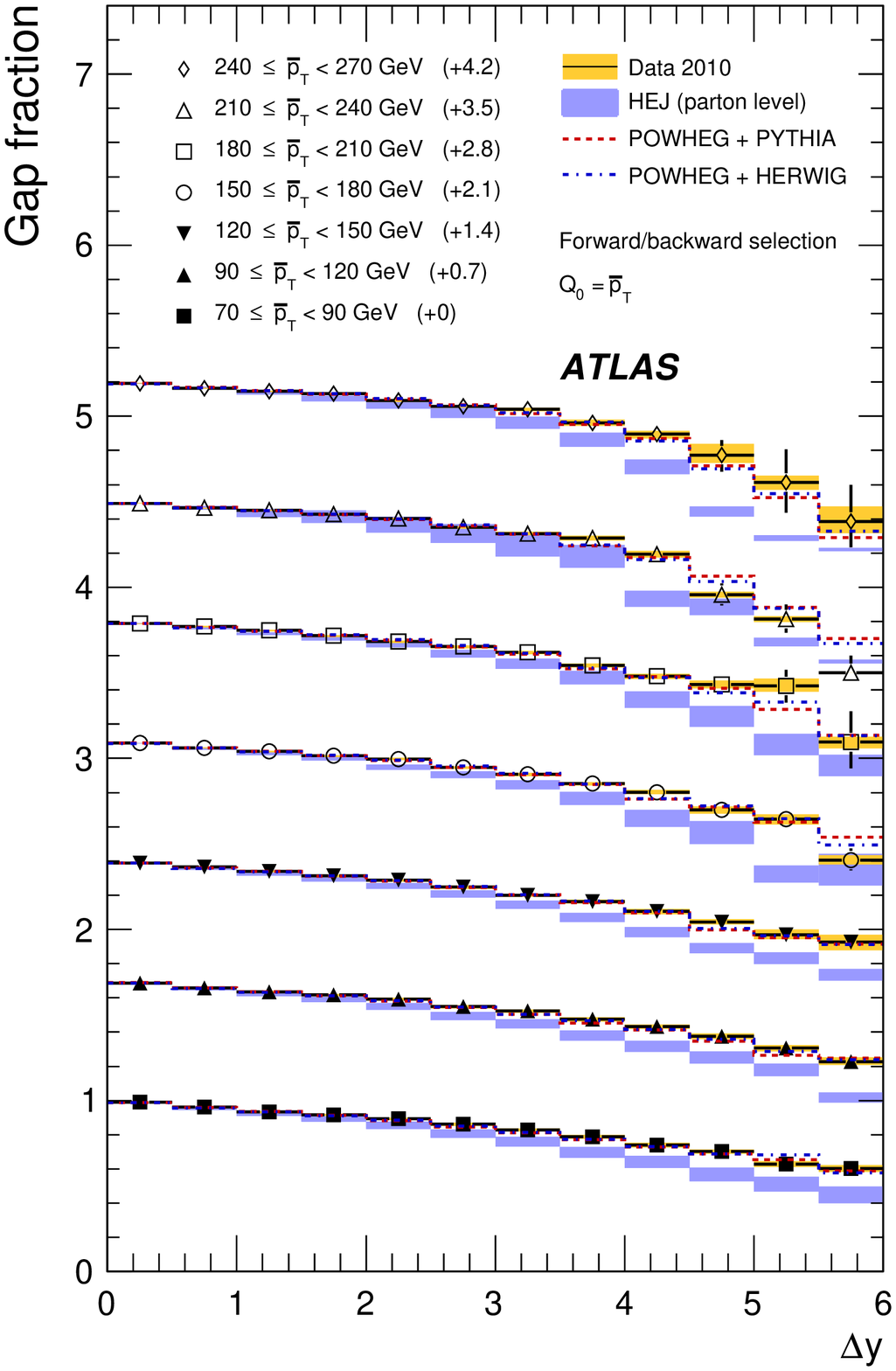}}
  \subfigure[]{ \includegraphics[width=.5\textwidth]{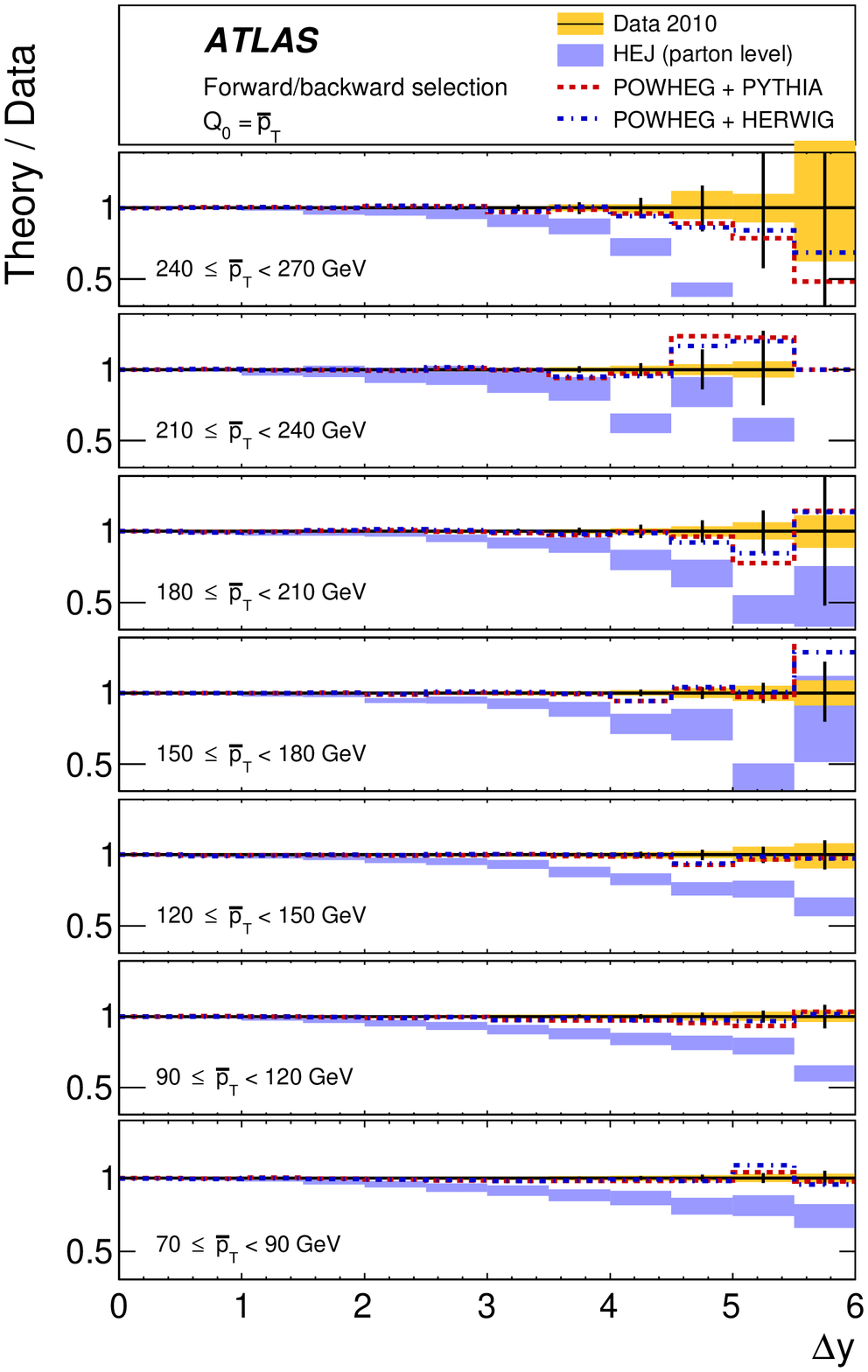}}
  }  
\caption{Gap fraction as a function of  $\Delta y$ for various $\overline{p}_{\rm T}$ slices. The dijet system is defined as the most forward and the most backward jets in the event and the veto scale is set to $Q_{0}=\overline{p}_{\rm T}$. The data are compared to the \hej{} and \powheg{} predictions in (a). The ratio of these theory predictions to the data are shown in (b).  The data and theory are presented in the same way as Fig.~\ref{fig:gf_vs_ptbarA}. \label{fig:gf_vs_dyBp}}
\end{figure}

In Fig.~\ref{fig:gf_vs_dyBp}, the dijet system is again defined as the most forward and the most backward jets in the event, but the veto scale is now set to $Q_{0} = \overline{p}_{\rm T}$.  In this case, both \powhegpyt{} and \powhegher{} give a good description of the gap fraction as a function of $\Delta y$, implying a smaller dependence on the generator modelling of the parton shower, hadronisation and underlying event. The \hej{} description of the data, however, does not improve with the increase in veto scale. 

\section{Summary}
A central jet veto was used to study the fraction of events that do not contain hadronic activity in the rapidity interval bounded by a dijet system (gap fraction). The dijet system was identified in two ways: using the two leading transverse momentum jets in the event and, alternatively, using the most forward and most backward jets in the event. The first approach examines the effect of wide-angle soft gluon radiation for $p_{\rm T}$-ordered jet configurations, whereas the second favours very forward-backward configurations and, therefore, BFKL-like dynamics. In addition, the mean number of jets in the rapidity interval bounded by the dijet system was presented, as an alternative variable for studying perturbative QCD emission.

The gap fraction was studied as a function of the rapidity separation between the boundary jets, $\Delta y$, the mean transverse momentum of the boundary jets, $\overline{p}_{\rm T}$ and the jet veto scale, $Q_{0}$. The mean number of jets in the rapidity interval was studied as a function of $\overline{p}_{\rm T}$ and $\Delta y$. In all cases, the data were corrected for detector effects. The data show the expected behaviour of a reduction of gap events, or an increase in jet activity, for large values of $\overline{p}_{\rm T}$ or $\Delta y$ \cite{Forshaw:2009fz}. The \pythia{}, \herwigpp{} and \alpgen{} leading-order MC event generators were compared to the data.  It was observed that \pythia{} and \herwigpp{} gave the best description of the data as a function of $\overline{p}_{\rm T}$ and that \pythia{} gave the best description of the data as a function of $\Delta y$.  \alpgen{} did not describe the data well at large values of $\overline{p}_{\rm T}$ or $\Delta y$. 

The data were compared to the NLO-plus-parton-shower predictions provided by \powheg{} when interfaced to \pythia{} (tune AMBT1) or \herwig{} (tune AUET1). In general, \powhegpyt{} gave the best description of the data, with \powhegher{} predicting too much jet activity in the rapidity interval between the boundary jets. Both \powheg{} predictions result in too low a gap fraction at large values of $\Delta y$, implying that the fixed order plus parton shower approach does not contain higher order QCD effects  that become important as $\Delta y$ increases.

The data were also compared to the \hej{} resummation of wide-angle emissions of similar transverse momentum. A particularly striking feature is that the parton-level \hej{} prediction has too little jet activity and too large a gap fraction at large values of $\overline{p}_{\rm T}/Q_{0}$. This means that the \hej{} calculation is missing higher order QCD effects that become important as $\overline{p}_{\rm T}/Q_{0}$ increases, i.e. those effects that are provided by a traditional parton shower approach. However, \hej{} does describe the data well as a function of $\Delta y$ when the dijet system is defined as the two leading $p_{\rm T}$ jets in the event and those jets do not have a value of  $\overline{p}_{\rm T}$ that is much larger than the veto scale. 

In most of the phase-space regions presented, the experimental uncertainty is smaller than the theoretical uncertainty.  Furthermore, the experimental uncertainty is much smaller than the spread of LO Monte Carlo event generator predictions. This data can therefore be used to constrain the event generator modelling of QCD radiation between widely separated jets. Such a constraint would be useful for the current Higgs-plus-two-jet searches and also for any future measurements that are sensitive to higher order QCD emissions.

\section*{Acknowledgements}

We thank Jeppe Andersen, Jeff Forshaw, Hendrik Hoeth, Frank Krauss, Simone Marzani, Paolo Nason, Emanuele Re, Mike Seymour and Jennifer Smillie for very useful discussions regarding the theory predictions used in this analysis.

We thank CERN for the very successful operation of the LHC, as well as the
support staff from our institutions without whom ATLAS could not be
operated efficiently.

We acknowledge the support of ANPCyT, Argentina; YerPhI, Armenia; ARC,
Australia; BMWF, Austria; ANAS, Azerbaijan; SSTC, Belarus; CNPq and FAPESP,
Brazil; NSERC, NRC and CFI, Canada; CERN; CONICYT, Chile; CAS, MOST and
NSFC, China; COLCIENCIAS, Colombia; MSMT CR, MPO CR and VSC CR, Czech
Republic; DNRF, DNSRC and Lundbeck Foundation, Denmark; ARTEMIS, European
Union; IN2P3-CNRS, CEA-DSM/IRFU, France; GNAS, Georgia; BMBF, DFG, HGF, MPG
and AvH Foundation, Germany; GSRT, Greece; ISF, MINERVA, GIF, DIP and
Benoziyo Center, Israel; INFN, Italy; MEXT and JSPS, Japan; CNRST, Morocco;
FOM and NWO, Netherlands; RCN, Norway; MNiSW, Poland; GRICES and FCT,
Portugal; MERYS (MECTS), Romania; MES of Russia and ROSATOM, Russian
Federation; JINR; MSTD, Serbia; MSSR, Slovakia; ARRS and MVZT, Slovenia;
DST/NRF, South Africa; MICINN, Spain; SRC and Wallenberg Foundation,
Sweden; SER, SNSF and Cantons of Bern and Geneva, Switzerland; NSC, Taiwan;
TAEK, Turkey; STFC, the Royal Society and Leverhulme Trust, United Kingdom;
DOE and NSF, United States of America.

The crucial computing support from all WLCG partners is acknowledged
gratefully, in particular from CERN and the ATLAS Tier-1 facilities at
TRIUMF (Canada), NDGF (Denmark, Norway, Sweden), CC-IN2P3 (France),
KIT/GridKA (Germany), INFN-CNAF (Italy), NL-T1 (Netherlands), PIC (Spain),
ASGC (Taiwan), RAL (UK) and BNL (USA) and in the Tier-2 facilities
worldwide.

\bibliographystyle{JHEP}
\bibliography{paper.bib}

\clearpage
\begin{flushleft}
{\Large The ATLAS Collaboration}

\bigskip

G.~Aad$^{\rm 48}$,
B.~Abbott$^{\rm 111}$,
J.~Abdallah$^{\rm 11}$,
A.A.~Abdelalim$^{\rm 49}$,
A.~Abdesselam$^{\rm 118}$,
O.~Abdinov$^{\rm 10}$,
B.~Abi$^{\rm 112}$,
M.~Abolins$^{\rm 88}$,
H.~Abramowicz$^{\rm 153}$,
H.~Abreu$^{\rm 115}$
% ...
\bigskip

% ...
$^{10}$ Institute of Physics, Azerbaijan Academy of Sciences, Baku, Azerbaijan\\
$^{11}$ Institut de F\'isica d'Altes Energies and Universitat Aut\`onoma  de Barcelona and ICREA, Barcelona, Spain\\
% ...
$^{48}$ Fakult\"{a}t f\"{u}r Mathematik und Physik, Albert-Ludwigs-Universit\"{a}t, Freiburg i.Br., Germany\\
$^{49}$ Section de Physique, Universit\'e de Gen\`eve, Geneva, Switzerland\\
% ...
$^{88}$ Department of Physics and Astronomy, Michigan State University, East Lansing MI, United States of America\\
% ...
$^{111}$ Homer L. Dodge Department of Physics and Astronomy, University of Oklahoma, Norman OK, United States of America\\
$^{112}$ Department of Physics, Oklahoma State University, Stillwater OK, United States of America\\
% ...
$^{115}$ LAL, Univ. Paris-Sud and CNRS/IN2P3, Orsay, France\\
% ...
$^{118}$ Department of Physics, Oxford University, Oxford, United Kingdom\\
% ...
$^{153}$ Raymond and Beverly Sackler School of Physics and Astronomy, Tel Aviv University, Tel Aviv, Israel\\
% ...
\end{flushleft}


\end{document}